\documentclass[a4paper,11pt]{article}
\usepackage{jcappub} 
\usepackage{hyperref}
\usepackage{orcidlink}
\usepackage{amsmath}
\usepackage{booktabs}
\usepackage{}
\usepackage[normalem]{ulem}

\title{ \boldmath A Tale of Two Couplings: Bayesian Selection in the Interacting Dark Sector}

\author[a,b]{Miguel Barroso Varela \orcidlink{0009-0006-9844-7661}}
\author[c,d]{Álvaro de la Cruz-Dombriz \orcidlink{0000-0002-7072-9396}}
\affiliation[a]{{Departamento de Física e Astronomia, Faculdade de Ciências, Universidade do Porto, Rua do Campo Alegre s/n, 4169-007 Porto, Portugal}}
\affiliation[b]{{Centro de Física das Universidades do Minho e do Porto, Rua do Campo Alegre s/n, 4169-007 Porto, Portugal}}
\affiliation[c]{{Departamento de Física Fundamental, Universidad de Salamanca, 37008 Salamanca, Spain}}
\affiliation[d]{{Cosmology and Gravity Group, Department of Mathematics and Applied Mathematics, University of Cape Town, Rondebosch 7700, Cape Town, South Africa}}

\emailAdd{up201907272@edu.fc.up.pt}
\emailAdd{alvaro.dombriz@usal.es}

\abstract{$\Lambda$CDM provides a remarkably successful description of the Universe, yet it suffers from persistent discrepancies, with one of the most prominent being the $\sigma_8$ tension between early-Universe Cosmic Microwave Background (CMB) measurements and late-time structure growth probes. To address this mismatch, we investigate generalized interacting dark sector models governed by a quintessence field coupled to dark matter via conformal and disformal transformations. By generically engineering a quintessence potential that ensures the combined evolution of dark matter and dark energy exactly mimics $\Lambda$CDM, our models satisfy all background constraints while signaling modified gravity effects within the linear perturbation sector. We perform a statistical analysis using late-time structure growth data anchored to Planck 2018 constraints. Furthermore, we address the initial conditions problem inherent to purely disformal models, demonstrating the necessity of a non-zero initial kinetic energy to initiate dynamical evolution. Our results indicate that purely conformal couplings efficiently transfer energy from dark matter to the quintessence field, suppressing late-time structure formation and successfully accommodating CMB data within late-time modified structure growth. This purely conformal limit yields a strong statistical preference over the $\Lambda$CDM baseline according to both the Akaike and Bayesian Information Criteria. Conversely, while non-zero disformal couplings introduce a kinetic friction that dampens the dark sector energy exchange, their expanded parameter space incurs heavy statistical penalties without providing a decisive improvement over the purely conformal scenario.
}

\begin{document}
\maketitle
\flushbottom

\section{Introduction}
\label{sec:intro}
The Concordance $\Lambda$CDM cosmological model provides a remarkably successful description of the Universe's cosmic   expansion history and the formation of large-scale structure through a simple underlying fundamental parameter space. However, the age of precision cosmology has brought to light several persistent discrepancies. A notable example is that of the $\sigma_8$ tension, which highlights an incoherence between the high level of matter clustering detected by the Cosmic Microwave Background (CMB) measurements from the early Universe achieved by the Planck collaboration \cite{Planck:2018vyg} and the smoother Universe observed by late-time probes like weak gravitational lensing and galaxy clustering \cite{Pezzotta:2016gbo,delaTorre:2016rxm,Shi:2017qpr}. The relative simplicity of the $\Lambda$CDM model can therefore be its downfall, as it lacks an evolution that is complex enough to explain this mismatch. 

To alleviate these tensions, Interacting Dark Energy (IDE) models have emerged as a class of theory of possible interest. At the phenomenological level, allowing an ener\-gy/momentum exchange between the least understood components of the cosmological framework, i.e., the dark sector composed of dark matter and dark energy, introduces additional dynamical degrees of freedom that can effectively alter the growth of structures while retaining the success of the background expansion at late times \cite{Yang:2018pej}. Therefore, given the low clustering observed near the present, models that suppress structure formation at late times through specific interactions in the dark sector may serve as a physically viable alternative to standard dynamical dark energy and stand as a promising resolution to the $\sigma_8$ tension \cite{BeltranJimenez:2021wbq}.

A theoretically motivated framework for modeling these interactions follows from scalar-tensor theories, where scalar fields couple to an effective geometry related to the metric seen by the remaining fields by conformal and disformal transformations. These couplings have been widely studied across both cosmology and high-energy physics \cite{Koivisto:2012za}. Outside of the cosmological context, conformal couplings frequently characterize theories of Axion-Like Particles (ALPs) and have been heavily constrained by high-precision photon fifth-force experiments \cite{Brax:2021wcv,Bauer:2024hfv,Grossman:2025cov}.
Conversely, disformal couplings, which depend upon the gradients and kinetic terms of a scalar field \cite{Bekenstein:1992pj}, can easily evade traditional fifth-force constraints within static non-relativistic environments due to their dependence on time-derivatives to be activated \cite{Koivisto:2012za}. Due to this, disformalities play an essential role in screening mechanisms and in the construction of string theory-inspired setups, such as the Dirac-Born-Infeld action describing the motion of D-branes in warped dimensions \cite{Sakstein:2014isa,vandeBruck:2015tna}. 

In this work, we propose a generalized interacting dark sector model with exponential field dependencies for both conformal and disformal couplings. In order to focus uniquely on the intricate dynamics of matter perturbations on the cosmological stage, we establish a quintessence potential that ensures that the combined evolution of coupled dark matter and dark energy mimics $\Lambda$CDM at the background level. This ensures the model satisfies the stringent background constraints from Baryon Acoustic Oscillations and Type Ia Supernovae, isolating cosmological modifications to the linear perturbation sector. We systematically categorize the parameter space and accurately test the theory after addressing the problem of initial conditions in purely disformal models, which we do by demonstrating that these require non-zero initial kinetic energy to initiate a dynamical evolution, unlike conformal models, which naturally drive the field from an initially ``frozen" state. 

Our analysis extends the original work on conformal and disformally coupled quintessence (DCQ) fields discussed in Refs. \cite{Barros:2018efl,Dusoye:2020wom,Dusoye:2021jne}. Specifically, Ref. \cite{Barros:2018efl} discussed the application of a conformally coupled dark sector to the $\sigma_8$ tension, focusing on fitting late-time $\sigma_8$ data. Alternatively, Ref. \cite{Dusoye:2020wom} investigated the cosmic expansion history of a generalization of the conformal model to include disformal couplings, while \cite{Dusoye:2021jne} focused on the effects of the same model on perturbative cosmological dynamics, without conducting a fit to data. In this paper, we generalize all of the former contributions into a consistent catalog of background and perturbative effects, later applying this theoretical and numerical foundation to conduct a thorough analysis of the statistical adequacy of the different classes of models to the latest structure growth data, which has grown both in quantity and quality since the release of the aforementioned papers.

This paper is organized as follows. In Section \ref{sec:theory}, we introduce the model's action and detail how the $\Lambda$CDM cosmic expansion history can be replicated by fixing parts of the theory without specifying a concrete potential for the quintessence field. We follow this by establishing the dimensionless dynamical system necessary to solve the background dynamics, together with a review of the effects of this modified theory at the perturbative level and a thorough discussion of the initial condition sensitivity of the phenomenological consequences of the model. Then, in Section \ref{sec:methodology}, we present the observational data used in our study, consisting of late-time $\{f\sigma_8,f,\sigma_8\}$ values, as well as the process of anchoring the theory to precise CMB measurements, and the detailed methodology used to obtain the statistical results. The systematic categorization of the parameter space into classes of theories and their theoretical limits are described in Section \ref{sec:models}. In Section \ref{sec:results}, we discuss the posteriors and maximum likelihoods obtained from the statistical analysis of each model, compare their quality of fit and determine the Bayesian model selection. We finalize in Section \ref{sec:conclusion}, where we declare our conclusions and propose future avenues for research on the investigated topic. We work in the $(-+++)$ metric convention and use units with $c=1$ and $\kappa^2=8\pi G=1$.

\section{Theoretical Framework}
\label{sec:theory}

\subsection{Action}
\label{subsec:Action_and_MimickingLCDM}
We consider an extension of the Einstein-Hilbert action with a metric describing a coupled geometry $\tilde g_{\mu\nu}$,  related to the decoupled metric $g_{\mu\nu}$ as
\begin{equation}
\label{relation_metrics}
\tilde{g}_{\mu\nu}=C(\phi)g_{\mu\nu}+D(\phi)\partial_\mu\phi\,\partial_\nu\phi\,,
\end{equation}
where $C(\phi)$ and $D(\phi)$ denote the conformal and disformal functions respectively, which are controlled by the quintessence field $\phi$.
The total action of the coupled quintessence model is written as
\begin{equation}
    S = \int {\rm d}^4x \sqrt{-g} \left[ \frac{R}{2} - \frac{1}{2}g^{\mu\nu}\partial_\mu\phi\partial_\nu\phi - V(\phi) \right] + S_b[g_{\mu\nu}] + S_{(c)}[\tilde{g}_{\mu\nu}^{}] \, ,
\end{equation}
where $R$ is the Ricci scalar and $g$ is the determinant of the non-tilded metric. The baryonic matter, contained in $S_b$, is taken to be decoupled, i.e., following metric  
$\{g_{\mu\nu}\}$ whereas the quintessence field $\phi$ is coupled to the evolution of a generic fluid $S_{(c)}$, the latter following a metric $\{\tilde{g}_{\mu\nu}\}$ and, as will be seen later, taken to represent dark matter. Throughout this work, we will refer to the frame defined by the gravitational metric $g_{\mu\nu}$ as the \textbf{decoupled frame}, as it governs the standard (baryonic) matter content. Conversely, the frame defined by the metric $\tilde{g}_{\mu\nu}$ will be referred to as the \textbf{coupled frame}, as it dictates the evolution of the dark matter fluid interacting with the quintessence field.

We consider an exponential field-dependence of the conformal and disformal coupling functions, which we write as
\begin{equation}
\label{couplings}
    C(\phi) = {\rm e}^{-2\alpha  \phi}    \quad  \quad \quad D(\phi) = D_m^4\, {\rm e}^{-2(\alpha+\beta)  \phi} \,,
\end{equation}
where $\alpha$,  $\beta$ and $D_m$ are constants, the latter accounting  for the inverse mass scale of the disformal coupling
and carrying dimensions of $[\text{mass}]^{-1}$. This functional form is particularly generic and elegant, since the background and perturbative cosmological dynamics will often depend on field derivatives of $\ln C$ or $\ln D$, which in this case reduce to simple constants. In more general scenarios, different fields could be expected to be coupled with different metrics, each with their own coupling functions and thus a set of parameters $\{\alpha_i,\beta_i,D_{m,i}\}$. However, since we focus solely on a model of quintessence coupled to dark matter, these three constants are the only additional parameters that are introduced. 

Beyond their mathematical elegance, these exponential parameterizations are strongly motivated by fundamental physics \cite{Brax:2019rwf,vandeBruck:2015tna,Koivisto:2013fta,Kaloper:2011jz}. For instance, in the context of string theory, the low-energy effective action naturally yields exponential conformal couplings via the dilaton field. Furthermore, the exponential disformal dependence is a generic feature of Dirac-Born-Infeld (DBI) frameworks describing the motion of D-branes in warped extra dimensions. In such setups, the scalar field represents the radial position of the brane, and the exponential factor reflects the geometry of a warped throat. From an Effective Field Theory (EFT) perspective, exponential dependencies as in \eqref{couplings} often arise as the leading-order terms in theories with softly broken shift symmetries, which are necessary to protect the quintessence field's mass from large radiative corrections.

By varying the action with respect to the metric $g_{\mu\nu}$, we obtain the field equations in the decoupled frame for the $g_{\mu\nu}$ metric 
\begin{equation}
\label{field_eqns}
    R_{\mu\nu}-\frac{1}{2}g_{\mu\nu}R=T_{\mu\nu}^{\phi}+T_{\mu\nu}^{b}+T_{\mu\nu}^{(c)}\,,
\end{equation} 
where $T_{\mu\nu}^{(c),b,\phi}$ denote the energy-momentum tensors of each of the exis\-ting fluids. Since we consider a coupled dark matter (denoted as ``$(c)$" from now onwards) and baryonic matter ($b$) to couple differently to the quintessence field, which is the reason why we separate their contributions to the evolution of the background geometry in the field equations.

Note that the stress-energy tensor for the coupled fluid in the decoupled frame of the $g_{\mu\nu}$ metric that appears in Eq. 
\eqref{field_eqns}
relates to that of the $\tilde g_{\mu\nu}$ metric by
\begin{eqnarray}
\label{eq:StressEnergy_MetricRelation}
T^{(c)}_{\mu\nu}=-\frac{2}{\sqrt{-g}}\frac{\delta S_{c}}{\delta g^{\mu\nu}}
=-\frac{2}{\sqrt{-\tilde{g}}}\sqrt{\frac{\tilde g}{g}}\,C(\phi)\frac{\delta S_{c}}{\delta\tilde g^{\mu\nu}}=C(\phi)\sqrt{\frac{\tilde g}{g}}\,\tilde T^{(c)}_{\mu\nu} \, ,
\end{eqnarray}
where $\tilde T^{(c)}_{\mu\nu}$ denotes the stress-energy tensor of the coupled fluid corresponding to the $\tilde g_{\mu\nu}$ metric. For a more detailed discussion of the relation between the stress-energy tensors at the background level in both the coupled and decoupled frames, we refer the reader to the Appendix \ref{app:Background}.

With this relation in mind, we find that the conservation equations are modified by the coupling between dark matter and quintessence \cite{Barros:2018efl}
\begin{equation}
    \nabla^\mu T_{\mu\nu}^{\phi}=-Q_0\partial_\nu\phi \quad \quad\quad \nabla^\mu T_{\mu\nu}^{c}=+Q_0\partial_\nu\phi \quad \quad\quad\nabla^\mu T_{\mu\nu}^{b}=0 \, ,
\end{equation}
such that the total stress-energy tensor is covariantly conserved, while there is also a flow of energy and momentum between the coupled quintessence and dark matter fields. This coupling at the level of energy-momentum conservation can be described in terms of an interaction term $Q_0$, defined in terms of the stress-energy tensor of the dark matter fluid as 
\begin{equation}
    Q_0=\frac{C_\phi}{2C}T^{c}+\frac{D_\phi}{2C}T_{\mu\nu}^{c}\,\partial^\mu\phi\,\partial^\nu\phi-\nabla^\mu\left(\frac{D}{C}T_{\mu\nu}^{c}\,\partial^\nu\phi\right)\, ,
\end{equation}
with $C_\phi$ and $D_\phi$ denoting the derivatives of $C(\phi)$ and $D(\phi)$ with respect to $\phi$, respectively. The trace of the tensor $T_{\mu\nu}^c$ in the decoupled frame is calculated using the decoupled metric as $T^c=T_{\mu\nu}^c g^{\mu\nu}$.

\subsection{Mimicking the \texorpdfstring{$\Lambda$}{L}CDM background}
Taking the spatially flat ($\Omega_k=0$) Friedmann-Lemaître-Robertson-Walker (FLRW) metric
\begin{equation}
\label{metric}
    {\rm d}s^2= g_{\mu\nu}{\rm d}x^{\mu}{\rm d}x^{\nu}= -{\rm d}t^2+a^2(t)\delta_{ij}{\rm d}x^i{\rm d}x^j
\end{equation}
as the background geometry, we recover the Friedmann equation 
\begin{equation}
    H^2=\frac{1}{3}(\rho_\phi+\rho_{c}+\rho_b)
\end{equation}
and the Raychaudhuri equation
\begin{equation}
\label{Raychaudhuri_eqn}
    \dot H=-\frac{1}{2}\left[(\rho_\phi+p_\phi)+\rho_{c}+\rho_b\right] \, ,
\end{equation}
where the dot denotes the derivative with respect to the cosmic time $t$. Note that the energy flow between the dark matter and quintessence fluids is irrelevant to the overall background dynamics, as any excess (or deficiency) of dark matter energy density will necessarily be compensated by the opposite effect in the density of quintessence. The latter field is taken to drive late-time acceleration. We have also taken the standard equation of state parameter for both baryons and dark matter ($w_{c}=w_b=0$). 

The assumption of the {\it coupled dark matter} behaving as a pressureless fluid, as seen in Eq. 
\eqref{Raychaudhuri_eqn}, is worthy of a careful discussion. As noted in Eq. \eqref{eq:StressEnergy_MetricRelation}, the stress-energy tensor in the coupled frame is non-trivially connected to its counterpart in the decoupled frame, being the latter in which we choose to work since it is the one governing all other matter content. Indeed, 
\begin{equation}
\label{eq:UpDown_StressEnergy}
\begin{aligned}
    T^{(c)\,\mu}_{
    \;\;\;\;\;\;\;\;\lambda} \equiv T^{(c)\,\mu\nu} g_{\nu\lambda} &= \sqrt{\frac{\tilde{g}}{g}}\,\tilde T^{(c)\,\mu\nu} \left( \tilde{g}_{\nu\lambda} - D(\phi)\,\partial_\nu\phi\,\partial_\lambda\phi \right)\\
    &=\sqrt{\frac{\tilde{g}}{g}} \left( \tilde T^{(c)\,^\mu}_{~~~~~\lambda} - D(\phi)\,\tilde T^{(c)\,\mu\nu} \partial_\nu\phi\, \partial_\lambda\phi \right)\, .
\end{aligned}
\end{equation}
Now, under the conditions of a homogeneous and isotropic background as per Eq. \eqref{metric}, we find
that the spatial components of the stress-energy tensor obey 
\begin{equation}\label{eq:Pressure_Coupled}
    T^{(c)\,i}_{~~~~~j} = \sqrt{\frac{\tilde{g}}{g}} \left(\tilde T^{(c)\,i}_{~~~~~j} - 0 \right)=0 \, ,
\end{equation}
where we have used the fact that $\partial_j\phi=0$ and $\tilde T^{(c)\,i}_{~~~~~j}=0$ for such a coupled fluid to behave as a perfect and pressureless matter in its own frame. Since this result is valid for all values of $i,j=1,2,3$, it implies that both the pressure and anisotropic stress of such a coupled fluid are null at the background level if they are equally null in the coupled frame. Also, resorting to Eq.
\eqref{eq:UpDown_StressEnergy} once again, 
one easily finds 
\begin{eqnarray}
T^{(c)\,0}_{~~~~~j} = 0\,,\;\; T^{(c)\,i}_{~~~~~0}=0\,,
\end{eqnarray}
for $i,j=1,2,3$.
Thus, in what concerns a homogeneous and isotropic background, the coupled dark matter behaves as a pressureless fluid in both frames (see the Appendix \ref{app:Background} for a more formal analysis). 

On the other hand, the density and pressure of dark energy are defined in terms of the quintessence field as
\begin{equation}
    \rho_\phi\equiv\frac{1}{2}\dot\phi^2+V(\phi) \quad \quad \quad P_\phi\equiv\frac{1}{2}\dot\phi^2-V(\phi) \, .
\end{equation}
Note that even if we consider a slowly-rolling field ($\dot\phi^2\ll V(\phi)$), such that we recover $\rho_\phi=-P_\phi$, this does not necessarily imply that $\dot\rho_\phi=0$, since in general $Q_0\neq0$. The dynamics of the field $\phi$ is determined by its conservation equation in the FLRW background
\begin{equation}
    \ddot\phi+3H\dot\phi+\partial_\phi V=+Q_0 \, ,
\end{equation}
where we see that the effect of $Q_0>0$ is to transfer energy from the dark matter fluid to the dark energy field.

In fact, if we wish to replicate a background evolution that mimics $\Lambda$CDM, we must impose a quintessence potential $V_\phi$ that dictates the dynamics of $\phi$ in such a way that the combination of dark matter and dark energy in this coupled model ($\rho_{c}+\rho_\phi$) replicates the combination of decoupled dark matter and a cosmological constant ($\rho_{\rm DM}+\Lambda$, where $\rho_{\rm DM}$ describes dark matter in the usual decoupled model). This ensures that our model inherently satisfies stringent geometric constraints from baryon acoustic oscillations and Type Ia supernovae, while localizing the modified gravity effects entirely within the perturbation sector.

This constraint is equivalent to imposing
\begin{eqnarray}
&&\rho_\phi=\Lambda+\rho_{\rm DM}-\rho_{c} \\
&& p_\phi=p_\Lambda=-\Lambda
\end{eqnarray}
or equivalently in terms of the field $\phi$
\begin{equation}
\dot\phi^2=\rho_\phi+p_\phi=\rho_{\rm DM}-\rho_{c} 
\end{equation}
and thus
\begin{equation}
\rho_\phi=\dot\phi^2+\Lambda \,.
\end{equation}
The necessary potential can therefore be determined as
\begin{equation}
    V(\phi)=\frac{1}{2}\dot\phi^2+\Lambda \, ,
\end{equation}
meaning that one can always in principle find a potential that generates a $\Lambda$CDM-like background through the coupled quintessence model.

\subsection{Background dynamics and dimensionless formulation}
\label{subsec:dimensionless_background}

To robustly evolve the background cosmology, we recast the Friedmann and continuity equations into an autonomous dynamical system. This dimensionless formulation maps the phy\-si\-cal parameter space into a compact domain, mitigating numerical instabilities during integration. We thus define the following dimensionless variables \cite{Dusoye:2020wom}
\begin{equation}
\label{defs}
    x^2 \equiv \frac{\dot{\phi}^2}{6H^2}, \quad y^2 \equiv \frac{V}{3H^2}, \quad z^2 \equiv \frac{\rho_c}{3H^2}, \quad u^2 \equiv \frac{\rho_b}{3H^2},
    \quad v^2 \equiv \frac{\rho_r}{3H^2},
    \quad \sigma \equiv \frac{D(\phi)}{C(\phi)} H^2 \quad \,.
\end{equation}
Here, $x, y, z,$ and $u$ correspond to the square roots of the fractional energy densities ($\Omega_X$) for the scalar field kinetic energy, scalar field potential, coupled fluid, and baryons, respectively. The variable $\sigma$ quantifies the dynamical strength of the disformal coupling. 

We define the conformal and disformal coupling parameters by the gradients of their respective coupling functions
\begin{equation}
    \lambda_C \equiv -\frac{C_{,\phi}}{C} = -2\alpha, \quad \lambda_D \equiv -\frac{D_{,\phi}}{D} = -2(\alpha + \beta) \quad  \, ,
\end{equation}
where we see that these are simply constants in the exponential parametrization \eqref{couplings} we have chosen for the coupled quintessence model.

The Friedmann constraint dictates that $x^2 + y^2 + z^2 + u^2 + v^2 = 1$, where in particular $v^2$ represents the fractional energy density of radiation, which is relevant to include in our analysis if we wish to determine the early-time behavior of the model in order to investigate its compatibility with the present stringent CMB data. The rate of change of the Hubble parameter is expressed as
\begin{equation}
    \hat{H} \equiv \frac{\dot{H}}{H^2} = -\frac{3}{2}\left(x^2 + z^2 + u^2\right) - 2v^2 \quad 
\end{equation}

Taking derivatives with respect to the e-folding number $N = \ln a$ and using the relevant conservation equations for each cosmological fluid component, the background evolution is governed by a closed system of ordinary differential equations \cite{Dusoye:2020wom}
\begin{equation}
\label{eq:dudN}
\frac{{\rm d}u}{{\rm d}N} = -\hat{H} u - \frac{3}{2} u\,, \quad 
\end{equation}
\begin{equation}
\label{eq:dxdN}
    \frac{{\rm d}x}{{\rm d}N} = -\hat{H} x - \frac{3}{2} x + \frac{\sqrt{3}}{2} \tilde{Q}_0 z^2\,, \quad 
\end{equation}
\begin{equation}
\label{eq:dzdN}
    \frac{{\rm d}z}{{\rm d}N} = -\hat{H} z - \frac{3}{2} z - \frac{\sqrt{3}}{2} \tilde{Q}_0 x z\,, \quad 
\end{equation}
\begin{equation}
\label{eq:dsdN}
    \frac{{\rm d}\sigma}{{\rm d}N} = \left[\sqrt{3}x(\lambda_C - \lambda_D) - 3(x^2 + z^2+u^2)-4v^2\right]\sigma \quad \,.
\end{equation}
Note that the energy exchange between the scalar field and the cold dark matter is governed by the normalized background interaction term, $\tilde{Q}_0$. For the assumed functional forms of $C(\phi)$ and $D(\phi)$, this term evaluates to
\begin{equation}\label{eq:Tilde_Q0_Definition}
    \tilde{Q}_0\equiv \frac{Q_0}{3H^2z^2} = \frac{\lambda_C(1 - 6\sigma x^2) + 3\sigma x(x\lambda_D + \sqrt{3})}{2 - 6\sigma x^2 + 3\sigma z^2} \,.
\end{equation}
It is also worthwhile to analyze the equation for the evolution of $\sigma$ and its dependence on the difference between $\lambda_C$ and $\lambda_D$. In the chosen parametrization of this paper, we see that $\lambda_C-\lambda_D=2\beta$, which completely isolates the dependence of the theory on the disformal exponent $\beta$. This is a consequence of $\sigma\propto D_m^4\exp(-2\beta\phi)$ and motivates the particular choice of including $\alpha$ in both the conformal and disformal sectors of the physical geometry metric $\tilde g_{\mu\nu}$ to which dark matter is coupled.

The system \eqref{eq:dudN} - \eqref{eq:dsdN} demonstrates that while the overall expansion history $H(N)$ behaves identically to $\Lambda$CDM, the internal energy composition of the dark sector is continuously altered by the $\tilde{Q}_0$ interaction. Uncoupled species, such as baryons and radiation, evolve independently according to standard continuity equations, unperturbed by the scalar field dynamics.

\begin{figure}
    \centering
    \includegraphics[width=0.8\linewidth]{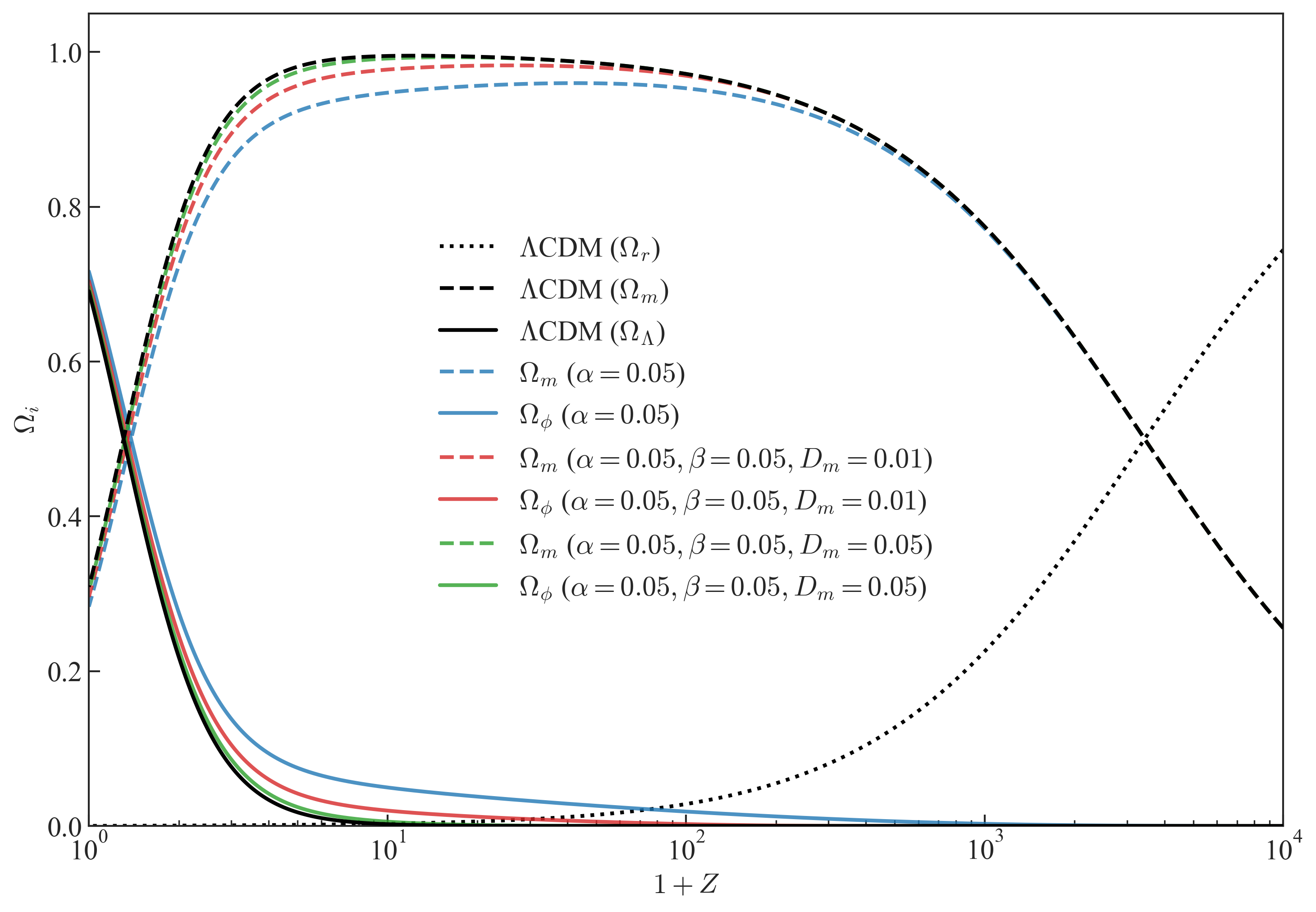}
    \caption{Evolution of the relative background densities of matter $\Omega_m=\Omega_b+\Omega_{c}$ and dark energy $\Omega_\phi$ in the standard $\Lambda$CDM model (black), a purely conformal model (blue), a disformal model (red) and a model with a constant disformal coupling (green). Values of $D_m$ are shown in units of meV$^{-1}$.}
    \label{fig:BackgroundDynamics}
\end{figure}

We show an example of the background dynamics in Fig. \ref{fig:BackgroundDynamics}. We have taken $\phi_i=\dot\phi_i=0$ as initial conditions for the quintessence field at a large redshift of $Z_i=10^5$, since the conformal coupling drives the field to gain kinetic energy and causes the matter, including baryons and the coupled fluid background, to deviate from $\Lambda$CDM standard evolution as the Universe expands. The radiation density is unaffected by the DCQ model, and given the delayed effect of the coupling on the dark matter fluid, the dynamics of recombination are safely maintained, as the detailed cosmological expansion history evolves exactly as in $\Lambda$CDM in the early Universe. The pure conformal model ($\alpha=0.05,\ D_m=0$) is shown in blue, where we see that in it both the matter and dark energy components exhibit the greatest deviation from $\Lambda$CDM. This is because the conformal coupling is able to drive the field's motion in an unrestrained fashion, thus allowing the deviation to become considerable.

In contrast, models with a non-zero  disformality ($D_m\neq0$), tend to stray less from the standard model. This can be easily interpreted by noting the association of $D(\phi)$ with a ``kinetic" term $\partial_\mu\phi\,\partial_\nu\phi$ in the metric $\tilde{g}_{\mu\nu}$, which leads to an additional friction term, re\-pre\-sen\-ted by $\sigma$ in the dimensionless first-order system discussed above. This suppresses the evolution of the field $\phi$, thus restraining the growth of the conformal coupling and therefore reducing the flow of energy between dark energy and dark matter through $Q_0$. In fact, it can be seen that this effect is highly sensitive to $D_m$, as in Fig. \ref{fig:BackgroundDynamics} we observe that, even when $D_m=\mathcal{O}(0.1 \ \text{meV}^{-1})$, the deviation from $\Lambda$CDM can become small enough to be almost negligible. This exact effect will play an important role in the discussion of different models of DCQ at the level of perturbations, where again $D_m$ will serve as a suppressor to the conformal sectors deviations from the standard model.

\subsection{Linear perturbations}
\label{subsec:perturbations}
As we take the model to perfectly reenact the background evolution of $\Lambda$CDM, its main distinction is through the perturbative sector of the theory. 

We start by considering scalar perturbations over the FLRW metric in the Newtonian gauge
\begin{equation}
\label{metric_FLRW_perturbed}
    {\rm d}s^2 = -(1+2\Phi){\rm d}t^2 + a^2(t)\delta_{ij}(1-2\Psi){\rm d}x^i {\rm d}x^j \,,
\end{equation}
where $\Psi$ and $\Phi$ are the usual Bardeen potentials. Additionally, we introduce perturbations to the matter content of the theory in the standard way:
\begin{equation}
\label{pert_Sec2}
    \rho_X(\vec{x},t) = \bar{\rho}_X(t) + \delta\rho_X(\vec{x},t), \quad P_X(\vec{x},t) = \bar{P}_X(t) + \delta P_X(\vec{x},t), \quad \phi(\vec{x},t) = \bar{\phi}(t) + \chi(\vec{x},t) \quad \, ,
\end{equation}
where $X=\{b,c,r\}$ denotes the fluid under consideration and quantities with a bar indicate the homogeneous background components of each fluid's density and pressure. We also define the density contrast $\delta_X\equiv\delta\rho_X/\bar\rho_X$ and the sound speed $c_{s,X}^2\equiv\delta P_X/\delta \rho_X$. 

To determine the sound speed of the coupled dark matter fluid, $c_{s,c}^{2} \equiv \delta P_{c}/\delta\rho_{c}$, we must evaluate the pressure perturbation $\delta P_{c}$ in the decoupled metric frame ($g_{\mu\nu}$). Rather than relying solely on the simplified background proportionality $P_{c} \propto \tilde{P}_{c}$, we must perturb the e\-xact spatial components of the stress-energy tensor established in Eq. \eqref{eq:UpDown_StressEnergy}.
Because the coupled dark matter behaves as pressureless dust in its own frame ($\tilde{g}_{\mu\nu}$), both its intrinsic background pressure and linear pressure perturbation vanish ($\tilde{P}_{c} = 0$, $\delta\tilde{P}_{c} = 0$). We must then consider the perturbative contribution of the disformal coupling term, $D(\phi)\tilde{T}^{(c)\,i\nu}\partial_{\nu}\phi\partial_{j}\phi$, appearing in Eq. \eqref{eq:UpDown_StressEnergy}. In a linearly perturbed FLRW universe, the background spatial field gradient is zero ($\partial_{j}\bar{\phi} = 0$). Consequently, the spatial derivative exists strictly at the perturbative level ($\partial_{j}\phi = \partial_{j}\chi$), making it a first-order quantity. Expanding the sum over $\nu$, the time component ($\nu = 0$) multiplies this first-order gradient by $(\tilde{T}^{c})^{i0}$, itself proportional to the spatial fluid velocity $\tilde{u}^{i}$, which is also a first-order quantity. Additionally, the spatial components ($\nu = k$) multiply it by another first-order spatial gradient $\partial_{k}\phi$. In all cases, the entire disformal contribution relies on the product of two perturbations, making it inherently of second order ($\mathcal{O}(\delta^2)$) or higher. Upon discarding these higher-order terms, consistent with our linear perturbation analysis, the pressure perturbation simplifies to the conformal scaling of the intrinsic fluid pressure
\begin{equation}
    \delta P_{(c)} = \delta \left( \sqrt{\frac{\tilde{g}}{g}} \right) \underbrace{\tilde{P}_{(c)}}_0 + \sqrt{\frac{\tilde{g}}{g}} \underbrace{\delta\tilde{P}_{(c)}}_0 = 0\,,
\end{equation}
thus implying that the sound speed of the coupled dark matter fluid is zero in the frame under consideration. Note that this also ensures that at leading order there is no anisotropic stress induced in the coupled fluid in this frame, since 
\begin{equation}
    (\delta T^{(c)})^{i}_j\sim\mathcal{O}(\delta^2)
\end{equation}
and thus the anisotropic stress in the decoupled frame is identically zero. For a detailed analysis of the relation between the components of the stress-energy tensor of the coupled fluid in the decoupled frame, see Appendix \ref{app:Perturbations}\footnote{Therein, we show in detail that the coupled fluid $i)$ is a pressureless fluid in both frames, and $ii)$ does not induce anisotropic stress in the non-tilded frame.}.

Since the total background dynamics is identical to those of $\Lambda$CDM and the baryonic content of the theory is trivially decoupled from all others, it is simple to find that the perturbations to the baryon density evolve according to an identical equation to the standard model
\begin{equation}\label{eq:Pert_Baryon_Diffeq}
    \delta_{b}'' + \delta_{b}'\left(2+\hat{H}\right) - \frac{3}{2}\left(\Omega_{b}\delta_{b} + \Omega_c\delta_c\right) = 0\,,
\end{equation}
where there will be a contribution from the DCQ modifications through the altered evolution of the dark matter content, present here in the form of both $\Omega_c$ and $\delta_c$. 

The evolution of the dark matter density perturbations is significantly more complex, since its coupling to the dynamical dark energy content of the theory extends to the perturbative sector in ways that inexorably links the two dark sectors. The detailed derivation of these equations is shown in Ref. \cite{Dusoye:2021jne}. For simplicity, we only show the final result here, which we write as
\begin{equation}\label{eq:Pert_DM_Diffeq}
\begin{aligned}
    \delta_c'' + \delta_c'(2 - 2\sqrt{3}\tilde{Q}_0 x + \hat{H}) &- \delta_c\left[\sqrt{3}\hat{q}_0 x + \frac{\tilde{Q}_0}{2}(7\sqrt{3}x + 3\tilde{Q}_0 z^2)\right] \\
    &+ \sqrt{3}\hat{q}_1 x + \frac{\tilde{Q}_1}{2}(7\sqrt{3}x - 3\tilde{Q}_0 z^2) - \frac{3}{2}(\Omega_c\delta_c + \Omega_{b}\delta_{b}) = 0 \,,
\end{aligned}
\end{equation}
where the definition of $\tilde Q_1\propto\delta_c$ and $\hat q_{0,1}$ is given in the aforementioned reference. Importantly, background quantities are denoted with a subscript ``$0$" while first-order perturbative quantities are denoted with a ``$1$", which clarifies that the above expression is indeed the leading-order contribution to the dynamics of the dark matter density perturbations. Note that we recover an identical equation to the baryonic sector of the theory in the limit where $\tilde Q_{0,1}=\hat q_{0,1}=0$, which is achieved by setting $C(\phi)=D(\phi)=0$, thereby restoring the trivial gravitational coupling in $\Lambda$CDM. 

\begin{figure}
    \centering
    \includegraphics[width=\linewidth]{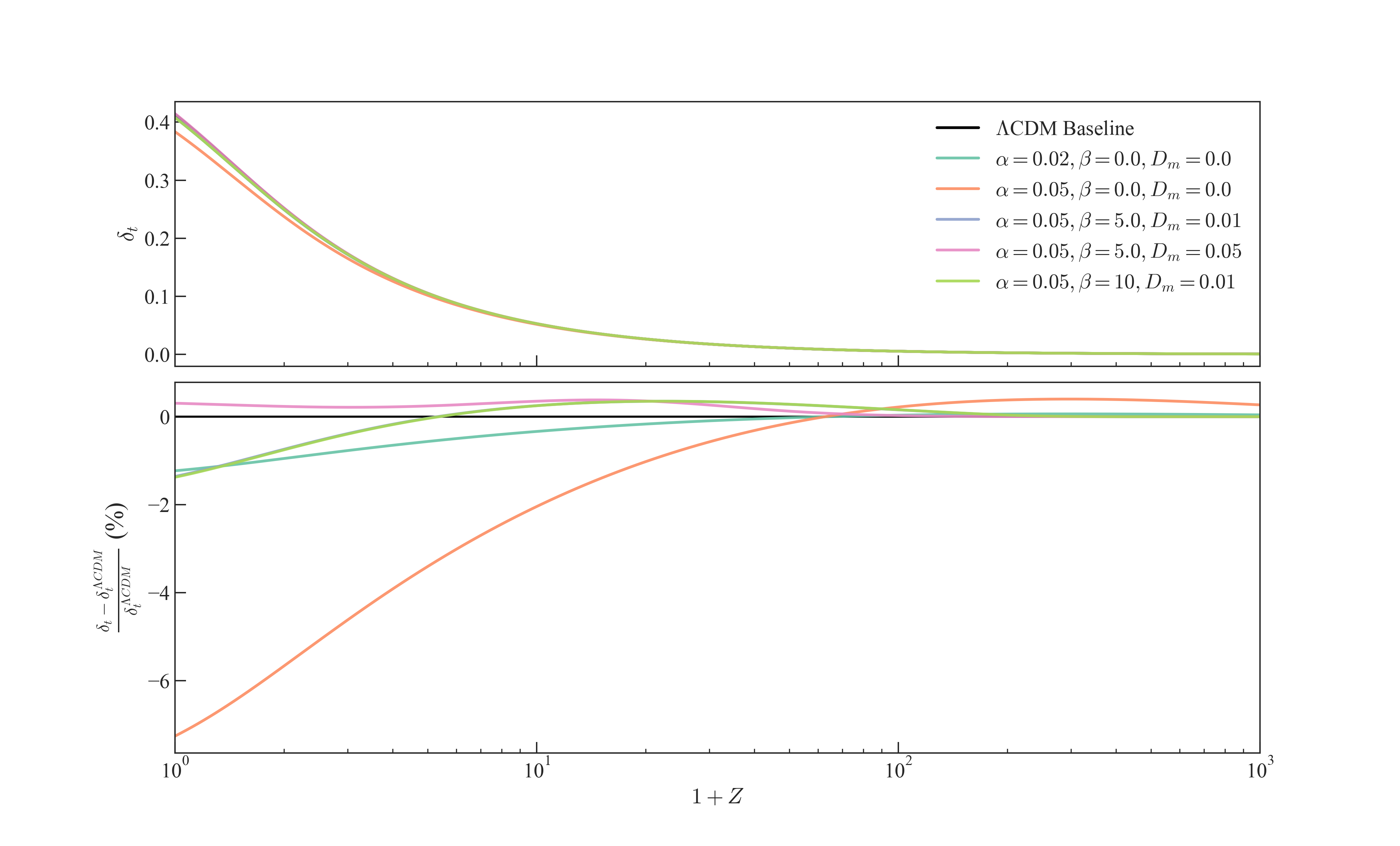}
    \caption{Top panel: The evolution of the total matter density contrast in the $\Lambda$CDM model (black) and in the DCQ model (colors). Bottom panel: The percentage difference of each model to $\Lambda$CDM. Values of $D_m$ are shown in units of meV$^{-1}$. For the considered values of $\alpha$ and $D_m$, curves differing only by different values of $\beta$ are indistinguishable and therefore totally overlap.}
    \label{fig:PerturbativeDynamics}
\end{figure}

When considering growth of large-scale structure throughout the Universe's evolution, we condense the contributions of baryonic and dark matter into a total matter density contrast defined as
\begin{equation}
    \delta_t \equiv \frac{\Omega_b\delta_b + \Omega_c\delta_c}{\Omega_b + \Omega_c} \,,
\end{equation}
which acts as a weighted average of both the baryonic and dark matter perturbations $\delta\rho_X\propto\Omega_X\delta_X$. We show examples of the effects of different DCQ models on this total contrast in Fig. \ref{fig:PerturbativeDynamics}. As seen at the background level in Fig. \ref{fig:BackgroundDynamics}, the effect of the conformal coupling is to remove energy from the dark matter sector and pass it onto the quintessence field, which is precisely what is observed at the perturbative level in Fig. \ref{fig:PerturbativeDynamics}. Due to this effect, purely conformal models ($D_m=0$) exhibit less clustering of matter content than in $\Lambda$CDM, as the loss of energy density to the dark energy sector counters the typical growth of structure. The leading culprit is the decrease in the dark matter background density $\Omega_c$, which is the main source of clustering in Eqs. \eqref{eq:Pert_Baryon_Diffeq} and \eqref{eq:Pert_DM_Diffeq}. This effect is controlled by the magnitude of $\alpha$, since the latter controls the rate at which the energy exchange in the dark sector occurs. However, at early times ($Z\gtrsim\mathcal{O}(10^3)$) the effect is the exact opposite, since at that epoch the background the evolution of the different components is still relatively similar to $\Lambda$CDM, as seen at the level of the individual relative densities $\Omega_X$ in Fig. \ref{fig:BackgroundDynamics}. This means that the dominating difference from $\Lambda$CDM at the perturbative level will be the fifth force introduced by the conformal coupling, contained in terms such as $\tilde Q_0$ in Eq. \eqref{eq:Pert_DM_Diffeq}. Nevertheless, this is ultimately outlasted by the leading-order effect due to the decrease in the background dark matter density $\Omega_c$.

In contrast to purely conformal models, models with a disformal component ($D_m\neq0$) tend to preserve the initial increase in clustering. This is due to the effect of these models on the background level evolution of the quintessence field $\phi$, since the additional friction term from the kinetic-like term in the disformally coupled part of the metric impedes the acceleration of the field value and thus the energy transfer from the dark matter to the dark energy densities. This is precisely why we see the curves with $D_m\neq0$ in Fig. \ref{fig:PerturbativeDynamics} staying above $\Lambda$CDM for longer for larger values $D_m$. In fact, if $D_m$ is large enough, this can even overcome the effect of the conformal coupling from the early Universe until the present, leading to an overall stronger clustering even at $Z=0$, although this difference from the standard model is quite small, being fixed at the sub-percent level. 

One can also see that the value of $\beta$ is largely irrelevant in the perturbative dynamics, as it was at the background level. This is due to its suppression for small kinetic terms $\dot\phi^2\ll1$, as the ones required to have a viable cosmological evolution. This can be seen mathematically through its association with a $x^2$ term in the definition of $\tilde Q_0$ in Eq. \eqref{eq:Tilde_Q0_Definition}. Given that $x\ll1$ for the duration of the evolution of the system, this is significantly sub-dominant to the principal effects from $\alpha$, which is itself associated with $\mathcal{O}(x^0)$  and $\mathcal{O}(x^1)$ terms, both of which dominate the $\mathcal{O}(x^2)$ coefficient of the disformal term following from $\beta$, i.e.~$\sigma$, in the same expression.

\subsection{The initial conditions problem and physical freeze-out}
\label{subsec:initial_conditions}

Before analyzing the theory against large-scale structure data, there is an important remark to be made about the initial conditions of the quintessence field $\phi$. If we intend on recovering a standard evolution of the theory at early times, so as to not spoil the finely tuned dynamics required for the emergence of temperature fluctuations in the CMB, which have been measured to remarkable precision in recent years, we must ensure the quintessence field is dormant in the initial stages of the Universe. A natural guess for a solution is to fix the field at its origin  ($\phi_i=0$), i.e. the field value at which the exponential dependencies of the conformal and disformal sectors are trivially 1, while ensuring it begins at rest ($\dot\phi_i=0$). This feels symmetric and can be simply achieved by engineering a dynamical quintessence potential which, although originally freezing the field at the origin, later undergoes some symmetry breaking process in the early Universe which frees $\phi$ to roll down and undergo the necessary evolution to build the $\Lambda$CDM background we observe today. Even in this limit of complete initial inactivity ($x_i\propto\dot\phi=0$), the field is brought to motion by its coupling to dark matter. This is easily seen by noting that
\begin{equation}
    \lim_{x_i \to 0} \tilde{Q}_0 = \frac{\lambda_C}{2}=-\alpha \, ,
\end{equation}
which, for $\alpha\neq0$, facilitates a flow of energy from dark matter to the quintessence field. 

However, when considering purely disformal models, i.e., those with no contribution from the conformal coupling ($\alpha=0$, $D_m\neq0$, $\beta\neq0$), this leads to a considerably important consequence. In this limit, we find 
\begin{equation}
    \lim_{\substack{x_i \to 0 \\ \alpha \to 0}} \tilde{Q}_0 = 0 \implies \lim_{\substack{x_i \to 0 \\ \alpha \to 0}} x'_i=0 \, ,
\end{equation}
such that $x=0$ for the entire duration of the system's evolution. This means that choosing an initially frozen field ($\dot\phi_i=0$) in purely disformal models leads to $\dot\phi=0$ and therefore $\phi=\phi_i$ for all phases of the Universe's evolution. Given that the contribution of the disformal sector follows from a kinetic-like term ($\partial_\nu\phi\,\partial_\mu\phi$), the absence of any kinetic energy of the quintessence field ensures that there is no modifications to the theory's dynamics, thus leading to trivial results that are completely identical to $\Lambda$CDM. We must therefore ensure the presence of some initial kinetic energy $\dot\phi_i\neq0$ in the case of purely disformal models, although without much loss of generality we can set $\dot\phi_i=0$ in any model with a non-trivial conformal sector ($\alpha\neq0$). One can think of such small non-zero field velocities as following from quantum fluctuations in the early Universe which are then boosted by the disformal sector to contribute significantly to modifications of the standard evolution of the Universe.

\section{Data and Statistical Methodology}
\label{sec:methodology}

Although the DCQ model under consideration is built with the restriction of reproducing a $\Lambda$CDM-like background and thus cannot be falsified or constrained through late-time data compilations that are only sensitive to the cosmic expansion history, such as Type Ia Supernovae and Baryon Acoustic Oscillations, its effects at the perturbative level leave unmistakable imprints on the growth of cosmic structures. In what follows, we introduce the methodology we will employ to constrain the model's parameters, while also determining its statistical preference to $\Lambda$CDM. 

\subsection{Observational datasets}
\label{subsec:data}
One of the most commonly used measurements in structure growth is the root mean square of matter fluctuations averaged over distances $8h^{-1}$ Mpc, labeled $\sigma_8$ and defined as
\begin{equation}
    \sigma_8^2(Z)=\int_0^\infty k^2\frac{{\rm d}k}{2\pi^2}P(k,Z)\,W^2(kR_8)\,,
\end{equation}
where $W(x)$ is the Fourier transform of a top-hat window function $W(x)=3(\sin x-x\cos x)/x^3$ and $R_8=8\,h^{-1}$ Mpc is the scale over which the matter fluctuations are averaged. As the matter power spectrum $P(k,Z)$ follows from a two-point correlation function of density perturbations, its amplitude evolves with the square of the growth factor $g(Z)$, defined as the ratio of the relative density contrast at some redshift $Z$ to its value at present, i.e., $g(Z)\equiv\delta(Z)/\delta(0)$, therefore being normalized as $g(0)=1$. Another useful quantity is the growth rate of the density perturbations, defined as $f\equiv\frac{{\rm d}\ln\delta}{{\rm d}\ln a}=\frac{\delta'}{\delta}$. In $\Lambda$CDM, the total ma\-tter perturbations evolve in a way that can be adequately parametrized by $f(Z)=[\Omega_m(Z)]^\gamma$, where $\Omega_m(Z)=H_0^2\Omega_{m,0}(1+z)^3/H^2$ is the relative matter density and the exponent has been constrained to be $\gamma\approx0.55$.

Since our model is scale-independent, meaning that the density perturbations at linear order are independent of the scale $k$ at which they are calculated, as can be seen in Eq. \eqref{eq:Pert_DM_Diffeq}, we can simply consider $\sigma_8=\sigma_{8,0}g(Z)$, where $\sigma_{8,0}$ is the value of $\sigma_8$ at present, such that the often measured quantity $f\sigma_8$ is proportional to the model-dependent quantity $f(Z)g(Z)$. In this work, we consider 66 points of $f\sigma_8$ data compiled in \cite{Skara:2019usd}, together with 12 points of isolated growth rate data compiled in \cite{Sahlu:2024dxp} and individual measurements $\sigma_8(Z)$ from \cite{Perenon:2019dpc}. As found in \cite{BarrosoVarela:2025zzv}, constraining alternative cosmological models requires a careful consideration of the correlation between the distinct data points. With this in mind, we include the explicitly calculated correlation of the WiggleZ, SDSS-IV and BOSS DR12 data, for which we use the independent covariance matrices provided in Ref. \cite{Sagredo:2018ahx}. We show all of the measurements included in the analysis of the DCQ model in Appendix \ref{sec:DataAppendix}.

\subsection{CMB anchoring and background parameters}
\label{subsec:cmb_prior}

Along the aforementioned late-time structure growth data, we restrict the early-time behavior of the theory in accordance with the constraints on $\Omega_{m,0}^{\rm Planck}=0.315\pm0.007$ and $\sigma_{8,0}^{\rm Planck}=0.811\pm0.006$ from the 2018 Planck results \cite{Planck:2018vyg}. These include the respective covariance matrix, which we also include, as there is a strong correlation between these quantities. However, since Planck constrains the value of $\sigma_{8,0}$ from the Cosmic Microwave Background at $Z\sim10^3$, we must extrapolate the DCQ model's prediction of $\sigma_8$ at this redshift before comparing it directly with the Planck value. As seen in Figure \ref{fig:PerturbativeDynamics}, we do not expect the model to play a dominant role at large redshifts, therefore allowing us to calculate the comparison value $\sigma_{8,0}^*$ as \cite{Perenon:2019dpc,BarrosoVarela:2025zzv}
\begin{equation}
\sigma_{8,0}^*=\sigma_{8,0}\frac{\delta_{\Lambda {\rm CDM}}(Z=0)}{\delta_{\rm DCQ}(Z=0)} \, ,
\end{equation}
where $\sigma_{8,0}$ is the free parameter that represents the real physical value of $\sigma_8$ at present in the DCQ model, which we include as a free parameter in the constraining process described in this section. The $\chi^2$ from the Planck measurements is thus calculated as
\begin{equation}
\chi^2_{\rm Planck}=\vec\Delta\cdot C^{-1}_{\rm Planck}\cdot\vec\Delta\, ,
\end{equation}
where 
\begin{equation}
    \vec\Delta=\left(\Omega_{m,0}-\Omega_{m,0}^{\rm Planck},\sigma_{8,0}^*-\sigma_{8,0}^{\rm Planck}\right)\,.
\end{equation}
Given the tight constraints (1-2\% precision) on these values from Planck, we expect these to tightly anchor the model's parameters to their CMB-derived values.

At this stage, there is an important point to stress about the interplay of early and late-time measurements of $\sigma_8$. In fact, there is an outstanding incoherence (the so-called ``$\sigma_8$ tension") between these methodologies. Late-time probes such as weak gravitational lensing and galaxy clustering from independent surveys observe a smoother universe with lower clustering values than those of the Planck collaboration, with differences at the $2\sigma$ to $3\sigma$ significance level in the $\Lambda$CDM model \cite{Joseph:2022jsf}. Although there is an open discussion in the literature regarding the merit of such a tension \cite{Abdalla:2022yfr}, there is undoubtedly a disjoint nature of these values which may be better explained by alternative cosmological models such as the one under study in this work.

In addition to the aforementioned free parameters, we also anchor the baryon and radiation densities to the Planck values $\Omega_{b,0}=0.049$ and $\Omega_{r,0}=9\times10^{-5}$ respectively \cite{Planck:2018vyg}. The radiation density is anchored to exquisite precision by measurements of the CMB temperature today and plays a phenomenologically negligible role in the late-time evolution and structure growth probed by our datasets. Similarly, the baryon density is tightly constrained by Big Bang Nucleosynthesis (BBN) and the relative amplitudes of the CMB acoustic peaks \cite{Planck:2018vyg}. Additionally, and perhaps most importantly, from a theoretical standpoint, the baryonic sector in our framework is explicitly decoupled from the quintessence field, evolving exactly as it does in standard cosmology (as dictated by its uncoupled conservation equation). Because the defining characteristics of the disformally coupled quintessence model are strictly confined to interactions within the dark sector, leaving $\Omega_{b,0}$ and $\Omega_{r,0}$ to flow independently would be an unnecessary expansion of the parameter space. This would only increase computational cost and risk diluting the constraining power of the data with uninformative degeneracies, without yielding any new physical insights into the coupled dark sector. Therefore, by constraining $\Omega_{m,0}$, we effectively constrain the coupled dark matter fluid's density $\Omega_{c,0}=\Omega_{m,0}-\Omega_{b,0}$, as well as the quintessence dark energy density $\Omega_{\phi,0}\approx1-\Omega_{m,0}$, which together make up the non-trivial sector of the modified theory.

\subsection{MCMC framework}
\label{subsec:mcmc}
We constrain the modified theory using a Markov Chain Monte Carlo (MCMC) sampler included in the \textsc{cobaya} package \cite{COBAYA_paper} and plot the resulting posteriors using the \textsc{GetDist} package \cite{Lewis:2019xzd}. We have calculated our likelihood function as $\mathcal{L}\propto\exp(-\chi^2/2)$, where the $\chi^2$ value of the full dataset is calculated as
\begin{equation}
\chi^2=\chi^2_{\sigma_8}+\chi^2_{f}+\chi^2_{f\sigma_8}+\chi^2_{\rm Planck}\,,
\end{equation}
with the corresponding covariance matrices included in its calculation when these are available and relevant. The chain convergence was calculated by the generalized version of the ($R-1$) Gelman-Rubin statistic that is built into \textsc{cobaya}, for which we kept the package's default criterion of $(R-1)=0.01$. 

Priors for the global parameters included in every model were taken to be wide and uniform, with $\Omega_{m,0}\in[0.1,0.5]$ and $\sigma_{8,0}\in[0.6,1.0]$, such that we introduce no bias towards late or early-time parameter preferences. For models including the conformal coupling parameter $\alpha$, we take its prior to be flat in the range $\alpha\in[0,0.5]$. A natural concern may arise when wondering why we do not open the model to negative values of $\alpha$. In fact, since $\alpha$ acts as the driving force for the quintessence field, displacing it from the ``frozen" state in which it initially finds itself in, models with opposite signs of $\alpha$ are completely identical apart from the direction in which $\phi$ evolves in field-space ($\alpha\rightarrow-\alpha\Leftrightarrow\phi\rightarrow-\phi$), which is irrelevant at the observational level, as originally discussed in Ref. \cite{Barros:2018efl}. Therefore, by including the decoupled regime ($\alpha=0$) and large positive values of $\alpha$ that lead to phenomenologically incompatible deviations from $\Lambda$CDM as the bounds in the uniform prior for this parameter, we ensure all plausible models are included in the MCMC analysis.

In what concerns models with a disformal component ($D(\phi)\neq0$), we will consider flat logarithmic priors in the disformality scale $\log_{10}D_m\in[-5,0]$, where we preserve our initial convention of presenting $D_m$ in units of meV$^{-1}$. Values as low as $D_m=10^{-5}$ meV$^{-1}$ effectively provide no disformality, while $D_m=1$ meV$^{-1}$ leads to a strong disformality that results in increased clustering in the late Universe (as seen in Fig. \ref{fig:PerturbativeDynamics} for $D_m$ as small as 0.01 meV$^{-1}$), which will be considerably disfavored by the lower clustering preferred by late-time data. The disformal exponent $\beta$ turns out to have little effect in the system's dynamics, as originally found in Refs. \cite{Dusoye:2020wom,Dusoye:2021jne} and discussed in the context of this work in Sec.  \ref{sec:theory}. Nevertheless, in the interest of generality, whenever $\beta$ is considered as a free parameter we utilize flat priors of $\beta\in[-10,10]$, in order to capture several orders of magnitude for this quantity, while also considering both positive and negative values, as this does make some difference in the context of its relative sign to $\alpha$, which can be thought of as setting the ``baseline sign" of the coupling coefficients in the exponent. In the case of the pure disformal model, we will also consider flat priors in the dimensionless field velocity $\log_{10}x_i\in[-5,-2]$, as values of $x_i$ that are too low lead to a recovery of the trivial $\Lambda$CDM model, while values that are too large lead to irreparable modifications to the early-time dynamics of the model.
As established earlier in this work, we fix our initial conditions at a large redshift of $Z_i=10^5$ to ensure the early-time behavior of all models is under control.

\subsection{Information criteria}
\label{subsec:information_criteria}

In order to classify the quality of the fit of each individual model with the aforementioned data, it is not enough to compare the minimum value of $\chi^2$ obtained from the MCMC analysis of each one. This is due to the varying number of free parameters present in each of these, which in this work ranges from a minimum of 2 -- as in $\{\sigma_{8,0},\Omega_{m,0}\}$ in $\Lambda$CDM -- to a maximum of 5, as for $\{\sigma_{8,0},\Omega_{m,0},\alpha,D_m,\beta\}$ in a general disformal model. 

With the interest of minimizing over-fitting, we present the results for two criteria for the quality of fit. The first of these is the Akaike Information Criterion (AIC), defined as
\begin{equation}
    {\rm AIC}=2k-2\ln\mathcal{L}^{\rm max}\,,
\end{equation}
where $k$ is the number of fitted parameters for each model and $\mathcal{L}^{\rm max}=-\chi^2_{\rm min}/2$ is the maximum posterior likelihood determined from the MCMC analysis \cite{AIC}. This criterion sets a linear penalty for models with a larger number of parameters, thus ensuring that a drop in the $\chi^2$ value must not be too small from the addition of a new parameter.

The second evaluation method is the Bayesian Information Criterion (BIC), which is defined as
\begin{equation}
    {\rm BIC}=k\ln N-2\ln \mathcal{L}^{\rm max}\, ,
\end{equation}
where $N$ symbolizes the number of data points in the observational sample analyzed \cite{BIC}. This criterion also has a linear penalty for the number of free model parameters, although the size of this penalty depends on the number of data points. For samples with $N\sim\mathcal{O}(1)$ this is practically equivalent to the AIC, while for the $N=83$ data points considered in the $\sigma_8+f+f\sigma_8+{\rm Planck}$ sample used in this work this penalty is approximately 2 times larger than the one for the AIC. We thus see that the BIC is a useful alternative to the AIC, since it prioritizes models with smaller numbers of fitted parameters, elevating the reward for simplicity in the structure of the theory and avoiding over-fitting. Following the criteria suggested in \cite{AIC_Evidence}, we establish that $\Delta {\rm AIC}>$ 2, 5 and 10, respectively, indicate weak, moderate, and strong evidence for the model with the higher AIC value, with an identical evaluation for the BIC.

\section{Parameter Space and Theoretical Limits}
\label{sec:models}
In this section, we briefly describe the assumptions behind each of the classes of models that we consider in this work. We present them in increasing order of complexity, as some of the models with additional free parameters include the simpler ones as a special case. We show a summary of all the combinations discussed in this section in Table \ref{tab:model_parameters}.

\begin{table}
\centering
\renewcommand{\arraystretch}{1.3}
\begin{tabular}{l c c c l}
\hline\hline
\textbf{Model} & $\boldsymbol{\alpha}$ & $\boldsymbol{\beta}$ & $\boldsymbol{D_m}$ & \textbf{Free Parameters} \\
\hline
$\Lambda$CDM Baseline & $0$ & - & $0$ & $\{\sigma_{8,0}, \Omega_{m,0}\}$ \\
Pure Conformal & $\neq 0$ & - & $0$ & $\{\sigma_{8,0}, \Omega_{m,0}, \alpha\}$ \\
Pure Disformal & $0$ & $\neq 0$ & $\neq 0$ & $\{\sigma_{8,0}, \Omega_{m,0}, D_m, \beta,x_i\}$ \\
Shift-Symmetric Limit & $\neq 0$ & $-\alpha$ & $\neq 0$ & $\{\sigma_{8,0}, \Omega_{m,0}, \alpha, D_m\}$ \\
Full Disformal Quintessence & $\neq 0$ & $\neq 0$ & $\neq 0$ & $\{\sigma_{8,0}, \Omega_{m,0}, \alpha, \beta, D_m\}$ \\
\hline\hline
\end{tabular}
\caption{Summary of the interacting dark sector models considered in the analysis, which include both conformally and disformally coupled quintessence models. The columns denote the constraints placed on the conformal ($\alpha$) and disformal ($\beta, D_m$) couplings, along with the resulting set of free parameters varied in the MCMC analysis. The pure disformal model requires a small, non-zero initial kinetic energy ($x_i \neq 0$) to initiate the quintessence field's evolution.}
\label{tab:model_parameters}
\end{table}

\subsection{The \texorpdfstring{$\Lambda$}{L}CDM baseline}
\label{subsec:lcdm}
Since we wish to determine the viability of the modified theory over the standard cosmological model, we will include $\Lambda$CDM as a baseline for our study. Naturally, this is achieved by setting  
\begin{equation}
    \alpha = 0 \quad \text{and} \quad D_m = 0\,,
\end{equation}
while the value of $\beta$ is made irrelevant as we remove the disformal sector through $D_m$. This model is fully described at the structure growth level through $\sigma_{8,0}$ and $\Omega_{m,0}$, with the former determining the level of clustering at present and the latter fixing the evolution of this clustering as we progress through the evolution of the Universe. It is easy to understand that this will by definition be the simplest model, having the minimal amount of free parameters. However, it is precisely this simplicity that leads to the rigidity between the early and late-time tension for $\sigma_8$, as the model is unable to explain the lack of clustering observed near the present in light of the stronger clustering predicted by CMB data.

\subsection{Pure conformal coupling}\label{subsec:conformal}

The simplest extension of the $\Lambda$CDM model within the interacting dark sector analyzed in this paper is the conformally coupled quintessence model. This was originally proposed in Ref. \cite{Barros:2018efl} under the context of structure growth, particularly motivated by the (then stronger) $\sigma_8$ tension. Its parameter space is simply defined by
\begin{equation}
    \alpha \neq 0 \quad \text{and} \quad D_m = 0 \,, 
\end{equation}
such that the disformal sector is completely removed, with the conformal sector completely determined by $\alpha$. This model reduces trivially to $\Lambda$CDM when $\alpha=0$, thus ensuring that the data are allowed to prefer the standard model through MCMC analysis. As previously discussed in Sec.  \ref{sec:theory}, the effect of $\alpha$ is to displace the quintessence field from its originally frozen state, gaining kinetic energy and therefore allowing for the interchange of dark energy with dark matter in the interacting dark sector through the conformal coupling. Since the dominant implication of this model is the reduction of the dark matter density at redshifts of order $Z\sim10-100$ (see Fig.  \ref{fig:BackgroundDynamics}) and a consequent decrease in the late-time clustering of matter (see Fig. \ref{fig:PerturbativeDynamics}), it is a promising alternative to the $\Lambda$CDM baseline model, as it may very well explain the decreased matter accumulation observed in late-time lensing and clustering data in comparison to early-time CMB data. 

\subsection{Pure disformal coupling}
\label{subsec:disformal}
Alternatively to the pure conformal model, one may instead disregard the conformal sector and focus entirely on the disformal coupling. This is defined by
\begin{equation}
    \alpha = 0, \quad \beta \neq 0, \quad D_m \neq 0 \quad\text{and}\quad x_i \neq 0 \, ,
\end{equation}
where we keep both $D_m$ and $\beta$ as free parameters, such that a constant disformal coupling is included as a special case when $\beta=0$. The standard $\Lambda$CDM model is recovered for $D_m=0$.

It is important to stress that for this to lead to any non-trivial consequences, we must ensure that $x_i\neq0$, i.e., the field must start with some non-zero (albeit small) kinetic energy. This follows from the lack of a ``driving force" when there is no conformally coupled sector in the theory. If one naively sets $x_i=0$, the purely disformal model is unable to start the movement of the field. This is seen by $\tilde Q_0$ being proportional to $x$ in Eq. \eqref{eq:Tilde_Q0_Definition} when $\lambda_C\propto\alpha=0$, while ${\rm d}x/{\rm d}N$ in Eq. \eqref{eq:dxdN} has terms proportional to $x$ and $\tilde Q_0$, meaning that $x_i=0\Rightarrow x'=0$ for all redshifts in pure disformal models, as discussed in Sec.  \ref{sec:theory}. The energy exchange in the dark sector is never kick-started and the model is trivially reduced to $\Lambda$CDM. To avoid over-fitting and the increase of computational cost, we investigate a discrete set of initial kinetic energies $x_i$, all constrained to be small ($x_i\lesssim10^{-4}$) by the requirement of recovering standard early-time background and perturbative dynamics. 

As expected, larger values of $D_m$ lead to stronger deviations from $\Lambda$CDM, with the associated increase in the scale of $\sigma\propto D(\phi)$ leading to an amplification of $\tilde Q_0$ in Eq. \eqref{eq:Tilde_Q0_Definition}. The sign of $\tilde Q_0$ is set by $x$ (since $\sigma>0$ by definition), although this sign is irrelevant since it is simply related to a change in the direction of the field's evolution. Importantly, its magnitude is proportional to $(\sqrt{3}+x\lambda_D)$, where for phenomenologically viable parameters we have a dominance of the constant $\sqrt{3}$ term. However, depending on the sign of $x\lambda_D$, this can lead to either a suppression or amplification of the interchange of energy between the quintessence and dark matter fluids. Since we fix a convention of $x>0$, this means that the sign of $\lambda_D\propto-\beta$ determines this effect. Although we do not show it here for the sake of brevity, we find that for $\beta>0$ we get a suppression of the interaction, while for $\beta<0$ we find the opposite.

\subsection{The shift-symmetric limit (\texorpdfstring{$\alpha = -\beta$}{alpha=-beta})}
\label{subsec:shift_symmetric}

The shift-symmetric limit represents a restricted case of the disformally coupled theory where we impose $\lambda_D=0$, which directly implies 
\begin{equation}
    \lambda_D = 0 \implies \alpha = -\beta\quad \text{and}\quad D_m \neq 0\,.
\end{equation}
In this specific scenario, the exponential field-dependence of the disformal coupling function vanishes entirely, yielding a constant disformal coupling $D(\phi) = D_m^4$. This model can be made purely disformal by setting $\alpha=0$, while $\Lambda$CDM is recovered by fixing both $\alpha=-\beta=0$ and $D_m=0$.
As discussed in Sec.  \ref{sec:theory}, the presence of a non-zero disformality introduces an additional friction term stemming from the kinetic-like contribution $\partial_\mu\phi\,\partial_\nu\phi$ in the metric $\tilde{g}_{\mu\nu}$. Because the disformal function is constant in this limit, this friction acts uniformly, suppressing the kinetic evolution of the quintessence field and restraining the growth of the conformal coupling without introducing additional dynamical field dependencies.  

This model serves as a highly physically motivated intermediate step, as it preserves the conformal driving force ($\alpha \neq 0$) necessary to naturally displace the field from its frozen initial state, while utilizing $D_m$ to damp the energy exchange in the dark sector and prevent excessive deviations from $\Lambda$CDM clustering. By fixing $\beta = -\alpha$, the parameter space is reduced by one dimension compared to the full disformal model. This effectively limits the model's vulnerability to the severe statistical penalties imposed by the BIC analysis, elevating the reward for theoretical simplicity while still testing the viability of a kinetic, field-independent disformal friction.

\subsection{Full disformal quintessence}
\label{subsec:full_model}

The full DCQ model represents the most general interacting dark sector scenario considered in this work, defined by 
\begin{equation}
    \alpha \neq 0, \quad \beta \neq 0 \quad \text{and}\quad D_m \neq 0\,.
\end{equation}
In this limit, both the conformal and disformal couplings are dynamically active and field-dependent, governed by their full exponential parametrizations presented in Eq. \eqref{couplings}.  This model possesses the maximum number of free parameters, expanding the fitting space to the full set of $\{\sigma_{8,0}, \Omega_{m,0}, \alpha, D_m, \beta\}$\footnote{As we will show in the remainder of the discussion of this paper, the posteriors are mostly degenerate in $\beta$, such that we can effectively treat it as a fixed parameter. However, this was not assumed a priori.}. Physically, this allows for a highly flexible interplay within the perturbative sector, with the conformal parameter $\alpha$ acting to transfer energy from dark matter to dark energy, lowering late-time matter density and mitigating structure clustering. Simultaneously, the disformal parameters $D_m$ and $\beta$ jointly dictate the magnitude and sign of the kinetic friction that opposes this exchange, allowing the model to uniquely preserve early-time clustering increases while suppressing them at late times. 

While this vast parameter space provides the highest theoretical potential to precisely fit the smoother universe observed in late-time data (potentially resolving the $\sigma_8$ tension), it inherently faces the strictest penalties from both the AIC and BIC. For this full model to be statistically preferred over the simpler purely conformal case or the $\Lambda$CDM baseline, the inclusion of a fully dynamical disformal sector must yield a substantial and decisive reduction in the overall $\chi^2$ capable of overcoming the heavy penalization for its five free parameters.

\section{Results}
\label{sec:results}

\subsection{\texorpdfstring{$\Lambda$}{L}CDM baseline constraints}
\label{subsec:results_lcdm}
\begin{figure}
    \centering
    \includegraphics[width=0.8\linewidth]{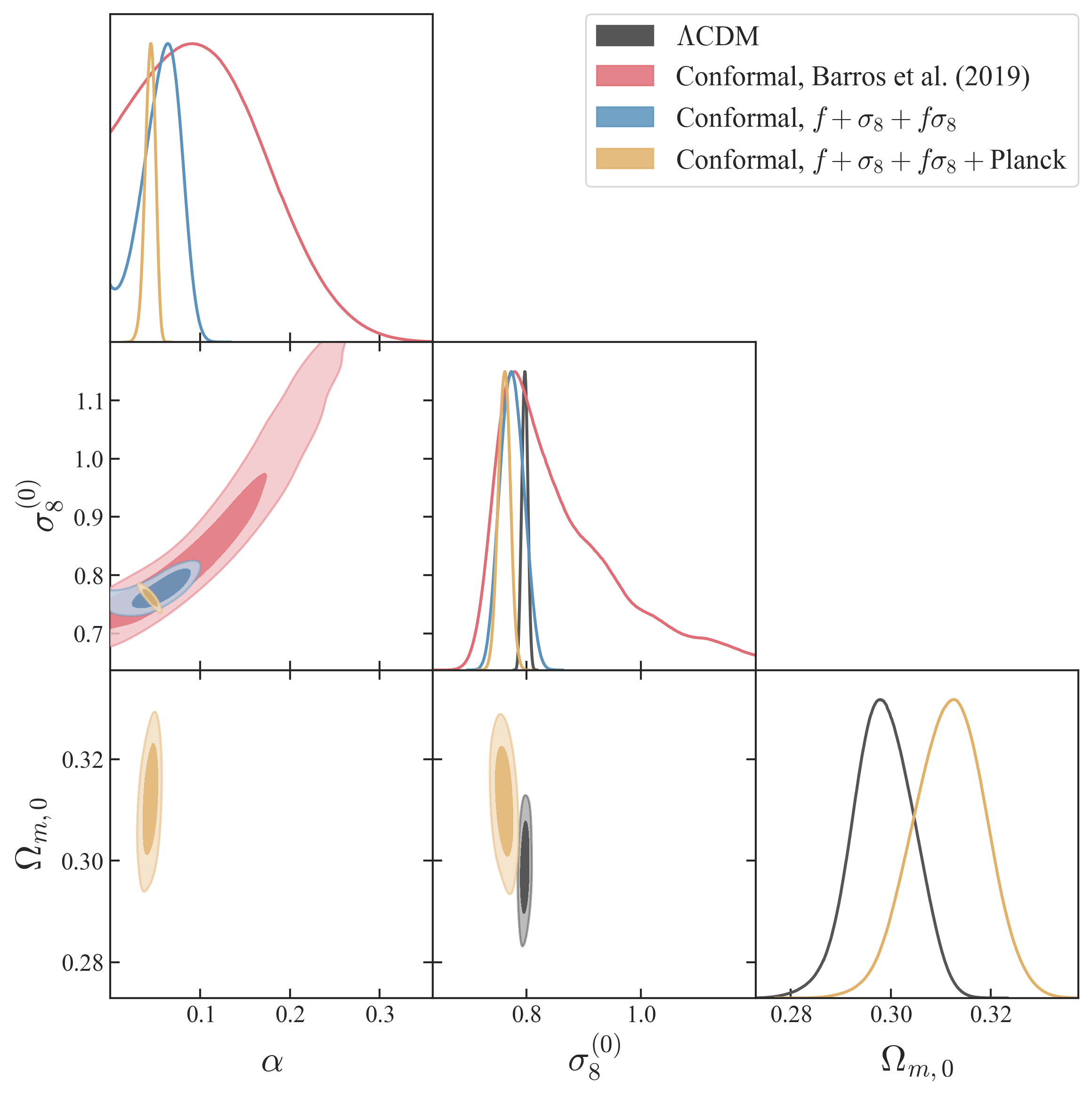}
    \caption{Posteriors for $\Lambda$CDM and the conformally coupled quintessence model. To illustrate the substantial improvement in the constraining power of recent large-scale structure data, we overlay the posteriors obtained using older datasets from the analogous analysis in Ref. \cite{Barros:2018efl}. The updated, significantly narrower posteriors demonstrate that the conformal model isolates a non-zero coupling ($\alpha \approx 0.045$), approximately 9 standard deviations from the $\Lambda$CDM limit. By yielding a reduced late-time clustering amplitude ($\sigma_{8,0} \approx 0.762$), the conformal model bridges the gap between Planck constraints and late-time structure measurements, considerably reducing the $\chi^2$ penalty compared to the standard model.}
    \label{fig:Conformal_Posteriors}
\end{figure}

We commence by presenting the results for the $\Lambda$CDM model, as this will serve as the baseline for the Bayesian model selection we will later conduct. Due to the rigidity of the model's evolution, reflected by its lack of parameters at the level of late-time structure growth, we find that the MCMC analysis quickly has to choose between the early or late-time data. We show the results for the posteriors of this model in Fig.  \ref{fig:Conformal_Posteriors}. Given the tight precision of the Planck constraints, the best-fit values $\sigma_{8,0}=0.798\pm0.005$ and $\Omega_{m,0}=0.299\pm0.006$ are closer to the Planck-derived measurements, although they are both approximately within $2\sigma$ of both the late and early-time datasets. However, this intermediate value does not escape the large $\chi^2$ penalty from both sides, with $\chi^2_{\rm min}=73.77$ being significantly higher than what we found when excluding the Planck data ($\chi^2_{\rm min}\sim58$), which we do not show here explicitly, as it was merely used as a comparison to assess the effect of the Planck data, as done in more detail in Ref. \cite{BarrosoVarela:2025zzv}.

This is not a surprising result, as it perfectly reflects the already established $\sigma_8$ tension, although it serves as a reinforcement of this incompatibility, which can only be broken by models with modified early or late-time dynamics at the perturbative level, such as the ones we discuss in what follows.

\subsection{Conformal coupling constraints}
\label{subsec:results_conformal}

We now focus on the results for the pure conformal model. This model extends the number of parameters of $\Lambda$CDM by one, as it adds the conformal coupling $\alpha$. As seen in Fig.  \ref{fig:PerturbativeDynamics}, this model leads to a slight over-clustering at $Z\sim100$, after which it quickly causes a stronger under-clustering, which can be principally attributed to the reduction in the relative matter density around $Z\sim10$, as exemplified in Fig.  \ref{fig:BackgroundDynamics}, which propagates towards the weaker density contrast near the present. 

This effect is precisely what is necessary to bridge between the larger value of $\sigma_8$ measured by Planck, and the lower value from direct late-time measurements. It effectively attributes the lower late-time value at the modifications of the conformally coupled dark matter, which ``leaks" energy to the quintessence field and therefore leads to a reduced growth of inhomogeneities. Because of this, the model is able to reach a much better statistical agreement with the full dataset used in this work. 

We show the corresponding posteriors in Fig. \ref{fig:Conformal_Posteriors}. We start by referring the original analysis of this model with $f\sigma_8$ data, which was conducted in Ref. \cite{Barros:2018efl}. In the 8 years since the release of that paper, there has been a considerable increase in the quantity and quality of structure growth data, as shown in the compilation of data we present in Appendix \ref{sec:DataAppendix}. In that work, the authors focused only on late-time $\sigma_8$ data, for which $\Lambda$CDM leads to a best-fit $\sigma_8=0.750\pm0.024$, while the conformal model shifts this to $\sigma_{8,0}=0.818^{+0.115}_{-0.088}$. Additionally, they found $\alpha = 0.110^{+0.043}_{-0.093}$, placing this parameter just a little over the $1\sigma$ significance level away from $\Lambda$CDM ($\alpha=0$).
While both models have similar values of $\chi^2\approx11$, the additional parameter introduced by the conformal model serves only the purpose of allowing more freedom for $\sigma_8$ and therefore significantly broadening its $1\sigma$ uncertainty to a level where it is well within the bounds imposed by Planck. 

Thanks to the great improvement in the data, the same analysis of only late-time data now allows us to constrain $\sigma_{8,0}=0.775\pm 0.020$, a significant refinement, as indicated by the narrower posteriors in Fig.  \ref{fig:Conformal_Posteriors}, which nevertheless remains well within the $1\sigma$ region of the posteriors from Ref. \cite{Barros:2018efl}. Additionally, we find $\alpha = 0.056^{+0.026}_{-0.017}$, which now places this parameter more than $3\sigma$ away from $\Lambda$CDM, even when just including late-time data. At the $\chi^2$ level, we find that this model ($\chi^2=53.5$) outperforms $\Lambda$CDM ($\chi^2=57.7$), placing both models in a practically equivalent footing according to the AIC and the BIC, proving the excellent adequacy of the conformal model in fitting late-time structure growth data. 

When including the Planck constraints on $\sigma_8$ and $\Omega_m$, which we now add as a free parameter for both models, the conformal models clearly outperform $\Lambda$CDM, as its added flexibility in the late stages of the Universe's evolution allows for better compa\-ti\-bi\-li\-ty between Planck and late-time data. The conformal model is thus able to accommodate a much lower $\sigma_{8,0}=0.762\pm 0.010$ (compared to $\Lambda$CDM's value of $\sigma_{8,0}=0.798$), as it achieves this through a shorter late-time clustering of matter. We also obtain $\Omega_{m,0} = 0.311\pm 0.007$, approximately $2\sigma$ away from the $\Lambda$CDM result of $\Omega_{m,0} =0.299$. The effect of the conformal model is particularly clear when assessing the conformal coupling parameter $\alpha=0.0446^{+0.0061}_{-0.0051}$, an astonishing $9\sigma$ distance from $\Lambda$CDM, while still within $3\sigma$ of the value obtained when analyzing only late-time data in this model. This shows how well the conformal model can accommodate the Planck measurements within the same regions of model parameter space as the late-time data. 

Finally, we note that the conformal model achieves all of this with a significantly lower value of $\chi^2=53.6$ than $\Lambda$CDM's $\chi^2=73.8$ in the presence of the Planck measurements. As we will discuss in more detail in Sec. \ref{subsec:model_selection}, both AIC and BIC tests will indicate strong evidence in favor of the conformally coupled quintessence model. In the context of the $\sigma_8$ tension, we should stress that the value of $\sigma_{8,0}=0.762\pm0.010$ obtained for this model is not a sign of a withstanding tension, since the model can accurately explain the Planck measurements even without raising the present value $\sigma_{8,0}$ to around 0.8.

\subsection{Disformal models}
\label{subsec:results_restricted}

Moving to the disformal quintessence scenarios, we first examine the restricted case where $\beta=0$. This retains the parameter space of the conformal model, apart from the addition of the disformal scale $D_m$. As discussed in Sec. \ref{sec:theory}, we expect the value of $\beta$ to be mainly irrelevant in the results of the theory. We have confirmed this at the posterior level, where we found that the posteriors in the parameter $\beta$ when $\alpha\neq0$ are mostly flat, and therefore fix $\beta=0$ as an exemplary case for the discussion that follows. This confirms the original claim from Ref. \cite{Dusoye:2021jne}, as well as our earlier theoretical predictions.

\begin{figure}
    \centering
    \includegraphics[width=0.8\linewidth]{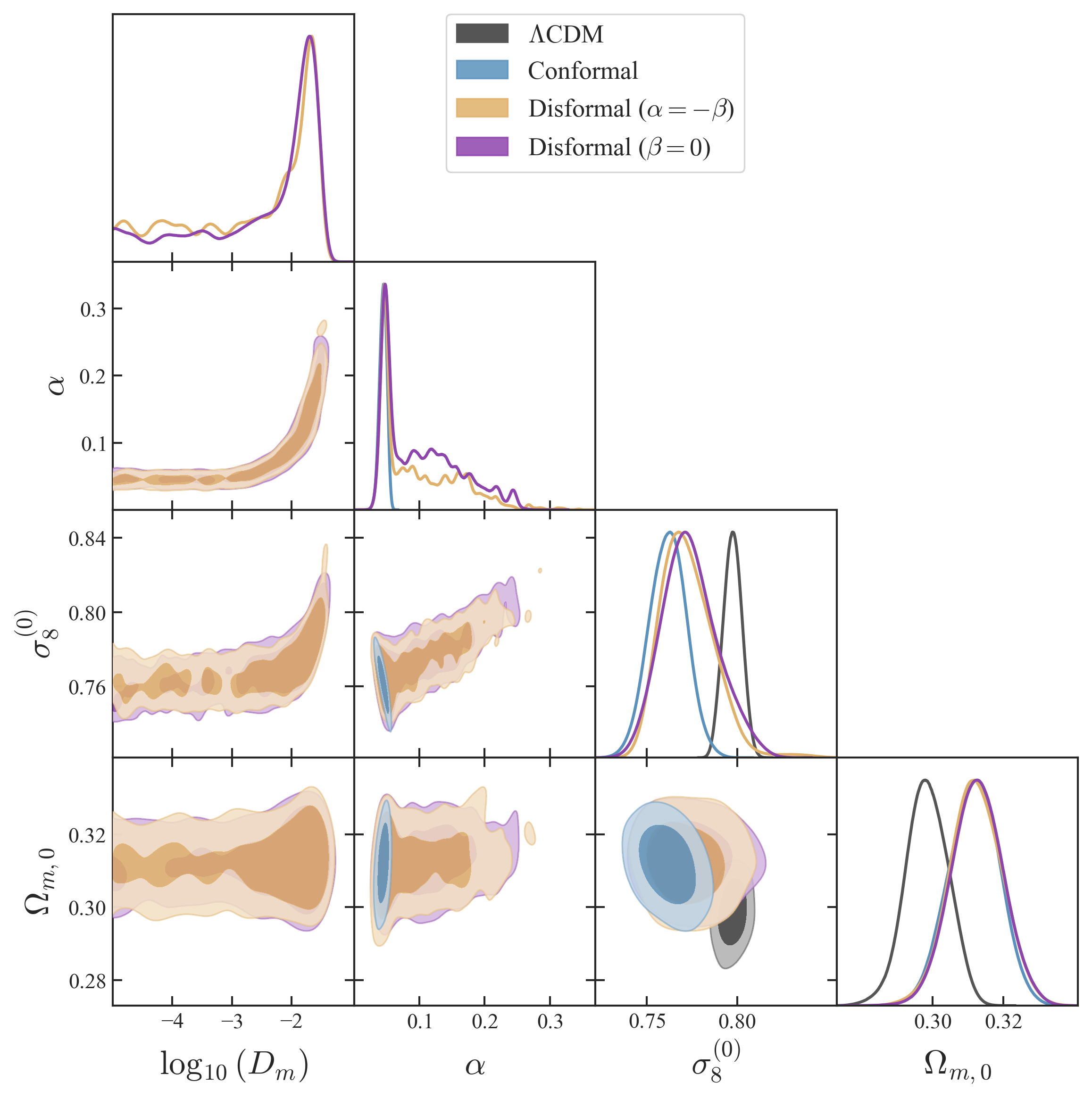}
    \caption{Posteriors for $\Lambda$CDM and the conformally and disformally coupled quintessence models. The inclusion of the disformal scale $D_m$ introduces an exponentially decaying dynamical friction that restricts the quintessence field's evolution. In the $\alpha-D_m$ plane, small disformal couplings ($D_m \lesssim 10^{-3}\text{ meV}^{-1}$) exert negligible friction, thus reducing the parameter space to the pure conformal limit ($\alpha \approx 0.05$). Conversely, larger $D_m$ values generate an ``L"-shaped degeneracy by permitting much stronger conformal couplings ($\alpha \lesssim 0.25$) before the friction ultimately decays and risks over-clustering. A detailed physical interpretation of these features in the parameter space is provided in Sec. \ref{subsec:results_restricted}.}
    \label{fig:Disformal_Posteriors}
\end{figure}

We show the results of this analysis in Fig. \ref{fig:Disformal_Posteriors}. Comparing the posteriors of this model and the ones for the conformal model above reveals some important insight on the principal difference between them. This can be seen in the additional freedom allowed by the disformal coupling parameter $D_m$. As already discussed in Sec. \ref{sec:theory}, the disformal sector introduces additional friction to the evolution of the quintessence field, thus reducing the rate of energy exchange within the dark sector of the theory. This effect is stronger with increasing $D_m$, which generates the ``L"-shaped posterior seen in the $\alpha-D_m$ panel of Fig. \ref{fig:Disformal_Posteriors}. As seen there, for small $D_m\lesssim10^{-3}\ \text{meV}^{-1}$ the effects of the disformal sector are so weak that the model behaves exactly like in the pure conformal limit, thus tightly funneling into the region with $\alpha\approx0.05$, as necessary to bridge between the early and late-time data. 

However, when moving into regions with larger $D_m\gtrsim10^{-3}\  \text{meV}^{-1}$, the suppression of the quintessence field's evolution allows for greater values of $\alpha$, with the posteriors extending up to $\alpha\lesssim0.25$ within the range $D_m\in[0.01,0.1]\ \text{meV}^{-1}$. Still, $D_m$ is not free to grow endlessly, as its effects can quickly lead to an over-clustering of matter, as seen in Fig. \ref{fig:PerturbativeDynamics}, which has the exact opposite effect of the conformal model and therefore only makes it harder to connect the results of Planck with those of late-time surveys. In this model, the disformal friction depends dynamically on the field itself, mathematically defined when $\beta=0$ as $D(\phi) = D_m^4 {\rm e}^{-2\alpha\phi}$.  An equivalent explanation is that, as the quintessence field evolves, this friction decays exponentially. Because the friction eventually fades away at late times, the model cannot support ever-increasing values of the driving force $\alpha$.

The situation is similar in the particular shift-symmetric case ($\alpha=-\beta$). The number of free parameters of this model is the same, as we fix $\beta$ for any given value of $\alpha$.  Ultimately, the addition of the disformal sector serves no purpose in facilitating compatibility with the data, as reflected by the unchanging minimum value of $\chi^2\approx53$, which overlaps with that of the conformal model. Given the addition of a free parameter ($D_m$), this will be disfavored by both AIC and BIC, with the latter punishing the additional parameter particularly strongly. This will be discussed in more detail in Sec. \ref{subsec:model_selection}.

\subsection{Purely disformal model}
\label{subsec:results_full} 

We now turn our focus to the purely disformal model ($\alpha=0$). As discussed in Secs. \ref{sec:theory} and \ref{sec:models}, this class has the particularity of requiring a non-zero initial field velocity $x_i$ to ``kick-start" the field's evolution, as there is no conformal sector to drive the field's dynamics initially. However, since this initial value determines the model's early-time behavior, which is stringently constrained by accurate CMB data, one cannot allow this value to become arbitrarily large.  In fact, we retain $x_i$ as a free parameter in order to remain as agnostic as possible.

Despite considering a broad parameter space, the purely disformal model proved unable to provide any satisfying fit to data. This follows from the tendency of the disformal model to retain the initial over-clustering caused by the coupling between dark matter and the quintessence field (see Fig. \ref{fig:Disformal_Posteriors}), which inevitably goes against the late-time under-clustering required to reconcile with the Planck data. The lack of insight provided by the posteriors of this model is why these have not been presented for the sake of brevity, as they reveal little to nothing of physical value. We thus exclude the purely disformal model from the following discussion of Bayesian model selection, as for all purposes its best fit can be considered as equivalent to $\Lambda$CDM, as this limiting case best describes the data within the restraints of a pure disformally coupled dark sector.

\subsection{Bayesian Model Selection}
\label{subsec:model_selection}

We now compare the different classes of models using both the AIC and the BIC. As noted above, the BIC penalizes additional free parameters more strongly than the AIC.
However, as we will shortly see, this will have no impact on the final conclusions on the statistical selection of models, as both criteria will point to the same qualitative results, even if with different quantitative values.

The AIC and BIC values for all of the investigated models (excluding the pure disformal models, as  justified above) are shown in Table \ref{tab:AIC_BIC}. Importantly, we present their relative values with respect to $\Lambda$CDM, which we treat as a baseline. Due to the inability of the standard model to explain the early and late-time data points simultaneously, its AIC and BIC values are considerably larger than that of the purely conformal model. We find $\Delta\text{AIC}=-18.19$ and $\Delta\text{BIC}=-15.76$, which indicate a strong preference for the modified theory over $\Lambda$CDM. This emphasizes the potential of models of conformally coupled quintessence to accurately describe the linearized growth of structure in the Universe while retaining a background evolution that perfectly mimics that of the current standard model.

As for models with a disformal sector, these were found to be equivalent at the level of fit quality, as determined by $\chi^2_{\rm min}$, therefore strongly favored over $\Lambda$CDM. However, due to their addition of an extra free parameter ($D_m$), they are disfavored relatively to the pure conformal model. In fact, the AIC points to weak evidence in favor of the conformal model, while the BIC is right on the boundary of moderate evidence.

\begin{table}
    \centering
    \renewcommand{\arraystretch}{1.3}
    \begin{tabular}{lcccc}
        \toprule
        \textbf{Model} & \textbf{AIC} & \textbf{BIC} & $\boldsymbol{\Delta}$\textbf{AIC} & $\boldsymbol{\Delta}$\textbf{BIC} \\
        \midrule
        $\Lambda$CDM         & 77.77 & 82.60 & 0 & 0 \\
        Pure Conformal                  & 59.58 & 66.84 & -18.19 & -15.76 \\
        Shift-Symmetric ($\alpha=-\beta$)         & 61.58 & 71.25 & -16.19 & -11.35 \\
        Full Disformal (Fixed $\beta$)  & 61.58 & 71.25 & -16.19 & -11.35 \\
        \bottomrule
\end{tabular}
    \caption{The AIC and BIC values for the models under study. We also present their individual comparison to the baseline case of the $\Lambda$CDM model. Models with lower AIC/BIC values are statistically favored. We exclude the pure disformal class of models, since their best-fit parameter space reduces to that of $\Lambda$CDM, although with worse AIC and BIC values due to the additional free parameters. Because $\beta$ is largely unconstrained/flat and fixed to $\beta=0$ in the full disformal runs, their minimum $\chi^2$ and active parameter count coincide.}
    \label{tab:AIC_BIC}
\end{table}

\section{Conclusions}
\label{sec:conclusion}
In this work, we investigated a generalized interacting dark sector model governed by a quintessence field coupled to dark matter via conformal and disformal transformations. By engineering a quintessence potential that ensures the combined background evolution of dark matter and dark energy exactly mimics $\Lambda$CDM, we are able to satisfy the stringent geometric background constraints while strictly isolating modified gravity effects within the linear perturbation sector. This construction provided a robust framework to address the persistent $\sigma_8$ tension between early Universe measurements and late-time structure growth probes.  

Prior to constraining the theory against observational datasets, we addressed the initial conditions problem inherent to purely disformal models. We demonstrated that a non-zero initial kinetic energy is strictly required to initiate dynamical evolution in these scenarios. Otherwise, the absence of a conformal driving force leaves the field frozen, trivially recovering standard $\Lambda$CDM dynamics.  

Our statistical analysis, utilizing late-time structure growth data anchored to Planck 2018 constraints, revealed that purely conformal couplings efficiently transfer energy from dark matter to the quintessence field. This continuous energy-momentum exchange suppresses late-time structure formation, naturally bridging the gap between the high clustering amplitudes predicted by CMB data and the smoother Universe observed by independent late-time surveys. Consequently, the purely conformal limit yields a strong statistical preference over the rigid $\Lambda$CDM baseline, as confirmed by substantial improvements in both the Akaike and Bayesian Information Criteria.  

Conversely, the inclusion of a disformal sector failed to yield a competitive statistical advantage. While non-zero disformal couplings introduce a kinetic friction that dynamically dampens the dark sector energy-momentum exchange, their expanded parameter space incurs heavy statistical penalties without providing a decisive improvement over the purely conformal scenario. Furthermore, purely disformal models exhibit an inability to relieve the $\sigma_8$ tension, as their inherent dynamics retain early-time over-clustering and fail to generate the necessary late-time suppression.  

Future avenues for research could explore the non-linear regime of structure formation within this coupled framework or investigate alternative functional forms for the conformal and disformal functions. Extending this analysis to include smaller scales will be essential to fully determine the theoretical viability of interacting dark sector scenarios as an alternative to the standard cosmological model.

\section*{Acknowledgments}The authors would like to thank Jose Beltr\'an Jim\'enez for insightful comments about the interpretation of these cosmological models.
The work of \textbf{MBV} is supported by FCT (Fundação para a Ciência e Tecnologia, Portugal) through the grant 2024.00457.BD and thanks the Erasmus+ scheme through the grant MIND PhD24 (2024-1-PT01-KA131-HED-000214341) for the support of the visit to the University of Salamanca, where part of this work was undertaken. Centro de Fisica do Porto is partially funded by Fundação para a Ciência e Tecnologia (FCT) under the grant UID04650-FCUP.
\textbf{AdlCD} acknowledges support from Project SA097P24 funded by Junta de Castilla y Le\'on (Spain), PID2024-158938NBI00 and CNS2024-154286 funded by MCIN/AEI/10.13039/\- 501100011033 and by {\it ERDF A way of making Europe}, and  NRF Grant CSUR23042798041 (South Africa).

\appendix

\section{Background Level Properties of the Coupled Fluid}\label{app:Background}

In this Appendix, we determine the properties of the coupled fluid in the decoupled frame at the background level, in the special case where this fluid can be treated as pressureless dark matter in the coupled frame, i.e., the frame in which this fluid is minimally coupled to gravity.

We start by defining 
\begin{equation}
\label{A_def}
A\equiv\sqrt{\frac{\tilde g}{g}}.
\end{equation}
where $\tilde g$ and $g$ denote the determinants of each metric, $\{\tilde{g}_{\mu\nu}\}$ and $\{g_{\mu\nu}\}$.

As mentioned above, we have assumed that in the coupled frame the coupled fluid is pressureless dust, therefore
\begin{equation}
\tilde T^{(c)\,\mu\nu}
=
\tilde\rho_c\,\tilde u^\mu\tilde u^\nu .
\label{T_tilde_coupled_munu}
\end{equation}
Consequently, at background level only the $00$ component, as measured by such a coupled observer, would be non-vanishing,
\begin{equation}
\tilde T^{(c)\,00}\neq0 .
\end{equation}
Using
\begin{equation}
\tilde T^{(c)\,\mu}_{\;\;\;\;\;\nu}
=
\tilde T^{(c)\,\mu\alpha}\tilde g_{\alpha\nu}\,,
\label{coupledTmunu_anyorder}
\end{equation}
which is valid at any perturbative order,
together with \eqref{A_def} and \eqref{eq:UpDown_StressEnergy},  the mixed tensor at the background level becomes
\begin{equation}
\label{coupledTmunu}
T^{(c)\mu}{}_{\nu}
=
A\,\tilde T^{(c)\,00}
\left(
\tilde g_{0\nu}\delta^\mu_{\;0}
-
D\,\partial_0\phi\,\partial_\nu\phi\,\delta^\mu_0
\right).
\end{equation}
Hence only the $\mu=0$ components may be non-zero, leaving $\nu=0,\,i$ $(i=1,2,3)$ as possible contributions. Thus, even if the coupled fluid is dust in the coupled frame ($\tilde{g}$), it is not in general in the decoupled frame ($g$).
More specifically, using the disformal relation 
\eqref{relation_metrics} at the background level for the 00 component,
\begin{equation}
\tilde g_{00}
=
Cg_{00}
+
D(\partial_0\phi)^2,
\end{equation}
we obtain
\begin{equation}
\tilde g_{00}
-
D(\partial_0\phi)^2
=
Cg_{00}.
\end{equation}
Similarly, since the metric $g$ - see \eqref{metric_FLRW_perturbed} - in the background satisfies $g_{0i}=0$, then
\begin{equation}
\tilde g_{0i}
=
\underbrace{D\,\partial_0\phi\,\partial_i\phi}_{\text{zero at background level}}.
\end{equation}
Therefore, since $g_{00}=-1$ in the background, at the background level the components of the mixed tensor satisfy
\begin{equation}
T^{(c)0}{}_{0}
=
A\,C\,g_{00}\tilde T^{(c)\,00},
\quad
T^{(c)0}{}_{i}=0,
\qquad
T^{(c)i}{}_{0}=0,
\qquad
T^{(c)i}{}_{j}=0\,
\label{T(c)0_0 background}
\end{equation}
since, at this level, $\phi$ is only a function of the zeroth variable ($t$). Recall that 
\eqref{coupledTmunu} is only valid at the background level. Also, as per the equations of motion written in the bulk of the article, we have defined 
the  {\it decoupled frame density} through
\begin{equation}
T^{(c)0}{}_{0}\equiv-\rho_c\,,
\label{rho_c_def}
\end{equation}
which through  
\eqref{T(c)0_0 background} 
would allow us to obtain $\tilde T^{(c)00}$ if need be.


\section{Perturbation Level Properties of the Coupled Fluid} \label{app:Perturbations}
Together with the metric scalar perturbations in the Newtonian gauge as per Eq. \eqref{metric_FLRW_perturbed}, we also need first-order perturbations on the scalar field as introduced in  \eqref{pert_Sec2}. Thus
\begin{equation}
\partial_0\phi
=
\dot{\bar\phi}+
\dot\chi
\,,
\qquad
\partial_i\phi
=
\partial_i \chi
\,,
\end{equation}
where let us remind that $\chi\equiv\chi(\vec{x},t)$ is already a first-order quantity. 
We can now proceed to calculate the components of the stress-energy tensor for the coupled fluid in the decoupled frame $T^{(c)\,\mu}_{\;\;\;\;\; \nu}
$ in terms of the equivalent quantity in the coupled frame $\tilde T^{(c)\,\mu}_{\;\;\;\;\; \nu}$.
\begin{itemize}
    \item $T^{(c)0}_{\;\;\;\;\;\ 0}$: Resorting to \eqref{eq:UpDown_StressEnergy}, for $\mu=\lambda=0$,
\begin{equation}
T^{(c)0}{}_{0}
=
A\left[
\tilde T^{(c)\,0}_{\;\;\;\;\;0}
-
D\,\tilde T^{(c)\,00}(\partial_0\phi)^2
\right].
\label{Exact_Tc00}
\end{equation}

Using
\begin{equation}
\tilde T^{(c)\,0}_{\;\;\;\;\;0}
=
\tilde T^{(c)\,00}\,\tilde g_{00},
\end{equation}
we obtain
\begin{equation}
T^{(c)0}{}_{0}
=
A\tilde T^{(c)\,00}
\left[
\tilde g_{00}
-
D(\partial_0\phi)^2
\right].
\end{equation}

Now use the disformal metric relation
\begin{equation}
\tilde g_{00}
=
Cg_{00}
+
D(\partial_0\phi)^2.
\end{equation}

The disformal contributions cancel exactly:
\begin{equation}
\tilde g_{00}
-
D(\partial_0\phi)^2
=
Cg_{00}.
\end{equation}

Hence the exact expression 
\eqref{Exact_Tc00} becomes
\begin{equation}
T^{(c)0}{}_{0}
=
ACg_{00}\,\tilde T^{(c)\,00}.
\label{Tc00 bis}
\end{equation}

Since at first order
\begin{equation}
g_{00}=-(1+2\Phi),
\end{equation}
Eq. \eqref{Tc00 bis} becomes
\begin{equation}
T^{(c)0}{}_{0}
=
-AC(1+2\Phi)\,\tilde T^{(c)\,00},
\label{A21}
\end{equation}
where $A$, $C$ and $(T^{(c)})^{00}$ above 
have not been expanded yet.
Should we now expand all quantities in \eqref{A21} to first order, namely
\begin{equation}
A=A_{\rm bg}+\delta A\,,\;\;
C=C_{\rm bg}+C_{{\rm bg},\phi}\,\delta\phi\,,
\end{equation}
\begin{equation}
\tilde T^{(c)\,00}
=
\tilde T^{(c)\,00}_{\rm bg}
+
\delta\tilde{T}^{(c)\,00}\,,
\end{equation}
where the subindex ${\rm bg}$  means background and ``$,\phi$" denotes the derivative with respect to $\phi$. Consequently, keeping only linear terms in \eqref{A21}, one gets
\begin{equation}
\begin{aligned}
T^{(c)0}{}_{0}
=
-A_{\rm bg}\,C_{\rm bg}\,\tilde T_{\rm bg}^{00}
\Bigg(
1
+
2\Phi
+
\frac{\delta A}{A_{\rm bg}}
+
\frac{C_{{\rm bg},\phi}}{C_{\rm bg}}\chi
+
\frac{\delta\tilde{T}^{(c)\,00}}{\tilde T_{\rm bg}^{00}}
\Bigg).
\end{aligned}
\label{Tc00_prefinal}
\end{equation}
Using the background definitions 
\eqref{T(c)0_0 background}
and \eqref{rho_c_def}, it is trivial to conclude that
\begin{equation}
\label{bar_rho_c} 
\bar\rho_c
=
A_{\rm bg}\,C_{\rm bg}\,\tilde T_{\rm bg}^{00},
\end{equation}
where both the overbar and the subscript `bg' denote background quantities throughout this Appendix. Finally, we obtain that \eqref{Tc00_prefinal} can be written as
\begin{equation}
T^{(c)0}{}_{0}
=
-\bar\rho_c
\left(
1
+
2\Phi
+
\frac{\delta A}{A_{\rm bg}}
+
\frac{C_{{\rm bg},\phi}}{C_{\rm bg}}\,\chi
+
\frac{\delta\tilde{T}^{(c)\,00}}{\tilde T_{\rm bg}^{00}}
\right).
\label{Tc00}
\end{equation}

Once the previous expressions have been presented, in the following we shall determine the relation between the density contrasts of the coupled fluid as defined in each frame. In order to do so, let us define
\begin{equation}
\gamma
\equiv
1-\frac{D_{\rm bg}}{C_{\rm bg}}\dot{\bar\phi}^{\,2}\,,
\end{equation}
with $\gamma$ obviously a background (zeroth-order) quantity.
Using
\begin{equation}
A=C^2\sqrt{1+\frac{D}{C}X},
\qquad
X=g^{\mu\nu}\partial_\mu\phi\,\partial_\nu\phi,
\end{equation}
which is valid at any order, and with
\begin{equation}
\bar X=-\dot{\bar\phi}^{\,2},
\end{equation}
we trivially conclude that 
\begin{equation}
A_{\rm bg}
\,=\,C_{\rm bg}^{\,2}\sqrt\gamma,
\end{equation}
Therefore, after a careful calculation, one finds the following first-order expression
\begin{equation}
\frac{\delta A}{A_{\rm bg}}
=
2\frac{C_{\rm bg,\phi}}{C_{\rm bg}}\chi
+
\frac{1}{2\gamma}
\left[
-\dot{\bar\phi}^{\,2}
\left(
\frac{D_{\rm bg,\phi}}{C_{\rm bg}}
-
\frac{D_{\rm bg}C_{\rm bg,\phi}}{C_{\rm bg}^2}
\right)\chi
+
\frac{2D_{\rm bg}}{C_{\rm bg}}
\left(
\Phi\,\dot{\bar\phi}^{\,2}
-
\dot{\bar\phi}\,\dot\chi
\right)
\right]\,.
\end{equation}

Defining the tilded density contrast as
\begin{equation}
\tilde\delta_c
\equiv
\frac{\delta\tilde\rho_c}{\bar{\tilde\rho}_c},
\end{equation}
and resorting to 
\eqref{T_tilde_coupled_munu} to obtain that
\begin{equation}
\tilde T^{(c)\,00}
=
\tilde\rho_c(\tilde u^0)^2\,,
\end{equation}
and using
\begin{equation}
\tilde g_{\mu\nu}\tilde u^\mu\tilde u^\nu=-1,
\end{equation}
one obtains
\begin{equation}
\frac{\delta\tilde{T}^{(c)\,00}}{\bar{\tilde T}^{00}}
=
\tilde\delta_c
+
\frac{1}{C_{\rm bg}\gamma}
\left[
-2C_{\rm bg}\Phi
-\left(
C_{{\rm bg},\phi}
-
D_{{\rm bg},\phi}\dot{\bar\phi}^{\,2}\right)\chi
+
2D_{\rm bg}\,
\dot{\bar\phi}\,\dot\chi
\right].
\end{equation}

Thus, by defining the density contrast for the coupled fluid in the untilded frame, $delta_c$, as follows
\begin{equation}
T^{(c)0}{}_{0}
=
-\bar\rho_c
\left(
1+\delta_c
\right),
\end{equation}
and resorting to Eq.  
\eqref{Tc00}, we can find that 
\begin{equation}
\delta_c
=
\tilde\delta_c
+
\left(1-\frac1\gamma\right)\Phi
+
\left[
\frac{5\gamma - 1}{2\gamma}\frac{C_\phi}{C}
+
\frac{D_\phi}{2C\gamma}\dot{\bar\phi}^{\,2}
\right]\chi
+
\frac{D}{C\gamma}\dot{\bar\phi}\,\dot\chi .
\end{equation}
which provides the relation between $\delta_c$ and $\tilde\delta_c$.
Therefore the first-order expression of the  (00) tensor component for the coupled fluid
energy-momentum tensor
in the untilded frame can be expressed as
\begin{equation}
T^{(c)0}{}_{0}
=
-\bar\rho_c
\Bigg[
1
+
\tilde\delta_c
+
\left(1-\frac1\gamma\right)\Phi
+
\left(
\frac{5\gamma - 1}{2\gamma}\frac{C_\phi}{C}
+
\frac{D_\phi}{2C\gamma}\dot{\bar\phi}^{\,2}
\right)\chi
+
\frac{D}{C\gamma}\dot{\bar\phi}\,\dot\chi
\Bigg].
\end{equation}

\item $T^{(c)0}_{\;\;\;\ \ i}$:
{ We explicitly evaluate the temporal-spatial ($0i$) components to illustrate below the emergence of the momentum exchange at the linear perturbation level. Resorting to the mixed tensor mapping, we extract the $\mu=0$, $\lambda=i$ component
\begin{equation}
    {T^{(c)0}}_{i} = A \left( \tilde{T}^{(c)\,0}_{\;\;\;\;\;i} - D(\phi)\tilde{T}^{(c)\,0\nu}\partial_{\nu}\phi\,\partial_{i}\phi \right)\,.
\end{equation}
Expanding the contraction above over the dummy index $\nu$ yields the separate time and spatial contributions,  specifically 
\begin{equation}
    {T^{(c)0}}_{i} = A \left[ \tilde{T}^{(c)\,0}_{\;\;\;\;\;\;i} - D(\phi) \left( \tilde{T}^{(c)\,00}\,\partial_{0}\phi + \tilde{T}^{(c)\,0j}\,\partial_{j}\phi \right) \partial_{i}\phi \right]\,.
    \label{T_c_0i}
\end{equation}
In a homogeneous and isotropic FLRW cosmology, the fluid possesses no background velocity, implying $\tilde{T}^{(c)0}_{\;\;\;\;\;\ i} = 0$. Also, given the fact that the scalar field exhibits no spatial gradients in the background, i.e., $\partial_{i}\bar{\phi} = 0$, the entire expression within the square brackets in Eq. \eqref{T_c_0i}  vanishes at the zeroth order. 

Now, to extract the first-order perturbation $\delta T^{(c)\,0}_{\;\;\;\;\; i}$, it can easily be shown that expression \eqref{T_c_0i} needs to be perturbed at first order, since generically $\delta T^\mu_{\ \ \nu}=\delta\left(T^\mu_{\ \ \nu}\right)$. Therein, because the pre-factor $A$ multiplies a term that is zero at the background level, any perturbation $\delta A$ evaluates to zero, leaving only the background value $A_{bg}$ to multiply the perturbed bracket. Within the square brackets of Eq. \eqref{T_c_0i}, the term that involves two spatial gradients, $D(\phi)\tilde{T}^{(c)\,0j}\,\partial_{j}\phi\,\partial_{i}\phi$, scales as $\mathcal{O}(\delta^3)$ and is discarded. This isolates the time-component of the disformal contraction as the sole eventual source of the first-order spatial gradient
\begin{equation}
    \delta T^{(c)\,0}_{\;\;\;\;\;\;i} = A_{\rm bg} \left[ \delta\tilde{T}^{(c)\,0}_{\;\;\;\;\;\;i} - D_{\rm bg}\,\tilde{T}^{(c)\,00}_{\rm bg}\,\dot{\bar{\phi}}\,\partial_{i}\chi \right]\,.
    \label{T_c_0i_pert}
\end{equation}
Recalling the definition \eqref{bar_rho_c} of the background dark matter density in the decoupled frame 
, we can express the tilded background energy density as $\tilde{T}^{(c)\,00}_{\rm bg} = \bar{\rho}_{c} / (A_{\rm bg} C_{\rm bg})$. Substituting this relation into \eqref{T_c_0i_pert} 
cancels the $A_{bg}$ term in the disformal piece, yielding the final simplified expression
\begin{equation}
    \delta T^{(c)\,0}_{\;\;\;\;\; i} = A_{\rm bg}\,\delta\tilde{T}^{(c)\,0}_{\;\;\;\;\;i} - \frac{D_{\rm bg}}{C_{\rm bg}}\bar{\rho}_{c}\,\dot{\bar{\phi}}\,\partial_{i}\chi\,.
  \label{T_c_0i_pert_final}
\end{equation}
This result demonstrates that even though coupled dark matter fundamentally behaves as pressureless dust with a standard linear momentum perturbation $\delta\tilde{T}^{(c)\,0}_{\;\;\;\; i}$ in its own (tilded) frame, the mapping to the decoupled (untilded) frame inevitably introduces a direct dependence on the scalar field's spatial gradient $\partial_{i}\chi$ provided $D_{\rm bg}\neq0$. This additional gradient term mathematically encodes the momentum exchange, manifesting physically as a ``fifth force", exerted by the quintessence field on the dark matter fluid.}
Note the fact that in our calculations the term $\delta T^{(c)\,0}_{\;\;\;\;\; i}$ has not been assumed to be null, rather encoding 
velocity perturbations, see for instance Ref.~\cite{Dusoye:2021jne} as well as \cite{Peter:2013avv, Ma:1995ey}. 
Thus, Eq. 
  \eqref{T_c_0i_pert_final} can be therefore just considered as a map relating 
$\delta T^{(c)\,0}_{\;\;\;\;\; i}$
and $\delta\tilde{T}^{(c)\,0}_{\;\;\;\;\; i}$
in case needed.

\item ${T^{(c)i}}_{j}$: 
We need to follow the same procedure for $i\neq j$ (a kind of anisotropic stress) or $i\equiv j$ (a kind of pressure) for $i,j=1,2,3$. Thus, resorting again to \eqref{coupledTmunu_anyorder}, which is valid at any order,
\begin{equation}
    {T^{(c)i}}_{j} = A \left[ \tilde{T}^{(c)\,i}_{\;\;\;\;j} - D(\phi)\,\tilde{T}^{(c)\,i\nu}\partial_{\nu}\phi\,\partial_{j}\phi \right]\,.
\end{equation}
Expanding the sum over the dummy index $\nu$ yields separate time and spatial contributions within the disformal contraction, in particular
\begin{equation}
    {T^{(c)i}}_{j} = A \left[ \tilde{T}^{(c)\,i}_{\;\;\;\;j} - D(\phi) \left( \tilde{T}^{(c)\,i0}\,\partial_{0}\phi + \tilde{T}^{(c)\,ik}\,\partial_{k}\phi \right) \partial_{j}\phi \right]\,.
\end{equation}
Following the same reasoning as above, in order to compute $\delta T^{(c)\,i}_{\;\;\;\; j}$ we know that 
\begin{enumerate}
    \item  $\tilde{T}^{(c)\,i}_{\;\;\;\;j}$ is null both in the background and perturbed level since $\tilde{T}^{(c)}$ is dust  and we are interested in such dust perturbations;
    \item  $\partial_j\phi$ only contributes at perturbation level, i.e., $\partial_j\phi = \partial_j\chi$;
    \item $\tilde{T}^{(c)\,ik}_{\rm bg}=0$ and $\tilde{T}^{(c)\,i0}_{\rm bg}=0$ since $\tilde{T}^{(c)}$ is dust.
\end{enumerate}
Thus we conclude
\begin{equation}
\delta T^{(c)\,i}_{\;\;\;\; j}=0
\end{equation}
for all possibilities of  $i,j=1,2,3$. In particular, the pressure perturbations $\delta T^{(c)\,i}_{\;\;\;\; i}$ in the decoupled (untilded) frame vanish entirely, a property that was explicitly utilized in the main body of this work. Similarly, the $i\neq j$ components of this perturbation being null serves as proof that there is no induced anisotropic stress on the coupled fluid in the decoupled frame.
\end{itemize}

\clearpage
\section{Data used in the analysis}
\label{sec:DataAppendix}

In this Appendix, we present the data used in the analysis of the coupled quintessence models against $f\sigma_8$ - combined with distinct $f$ and $\sigma_8$ datasets, see Table \ref{tab:combined_f_sigma} below - used in Sec. \ref{subsec:data}.

\begin{table*}[ht!]
    \centering
    \caption{$f\sigma_8$ measurements from Ref.\,\cite{Skara:2019usd}.}
    \label{tab:fsigma8}
    \setlength{\tabcolsep}{0pt} 
    \renewcommand{\arraystretch}{1.2} 
    \begin{tabular*}{\textwidth}{@{\extracolsep{\fill}} cc cc cc cc}
        \toprule
        $z$ & $f\sigma_8$ & $z$ & $f\sigma_8$ & $z$ & $f\sigma_8$ & $z$ & $f\sigma_8$ \\
        \midrule
        0.35 & $0.440\pm0.050$   & 0.77  & $0.490\pm0.180$  & 0.17  & $0.510\pm0.060$   & 0.02  & $0.314\pm0.048$   \\
        0.02 & $0.398\pm0.065$   & 0.25  & $0.3512\pm0.0583$& 0.37  & $0.4602\pm0.0378$ & 0.25  & $0.3665\pm0.0601$ \\
        0.37 & $0.4031\pm0.0586$ & 0.44  & $0.413\pm0.080$  & 0.60  & $0.390\pm0.063$   & 0.73  & $0.437\pm0.072$   \\
        0.067& $0.423\pm0.055$   & 0.30  & $0.407\pm0.055$  & 0.40  & $0.419\pm0.041$   & 0.50  & $0.427\pm0.043$   \\
        0.60 & $0.433\pm0.067$   & 0.80  & $0.470\pm0.080$  & 0.35  & $0.429\pm0.089$   & 0.18  & $0.360\pm0.090$   \\
        0.38 & $0.440\pm0.060$   & 0.32  & $0.384\pm0.095$  & 0.32  & $0.480\pm0.100$   & 0.57  & $0.417\pm0.045$   \\
        0.15 & $0.490\pm0.145$   & 0.10  & $0.370\pm0.130$  & 1.40  & $0.482\pm0.116$   & 0.59  & $0.488\pm0.060$   \\
        0.38 & $0.497\pm0.045$   & 0.51  & $0.458\pm0.038$  & 0.61  & $0.436\pm0.034$   & 0.38  & $0.477\pm0.051$   \\
        0.51 & $0.453\pm0.050$   & 0.61  & $0.410\pm0.044$  & 0.76  & $0.440\pm0.040$   & 1.05  & $0.280\pm0.080$   \\
        0.32 & $0.427\pm0.056$   & 0.57  & $0.426\pm0.029$  & 0.727 & $0.296\pm0.0765$ & 0.02  & $0.428\pm0.0465$ \\
        0.60 & $0.480\pm0.120$   & 0.86  & $0.480\pm0.100$  & 0.60  & $0.550\pm0.120$   & 0.86  & $0.400\pm0.110$   \\
        0.10 & $0.480\pm0.160$   & 0.001 & $0.505\pm0.085$  & 0.85  & $0.450\pm0.110$   & 0.31  & $0.469\pm0.098$   \\
        0.36 & $0.474\pm0.097$   & 0.40  & $0.473\pm0.086$  & 0.44  & $0.481\pm0.076$   & 0.48  & $0.482\pm0.067$   \\
        0.52 & $0.488\pm0.065$   & 0.56  & $0.482\pm0.067$  & 0.59  & $0.481\pm0.066$   & 0.64  & $0.486\pm0.070$   \\
        0.10 & $0.376\pm0.038$   & 1.52  & $0.420\pm0.076$  & 1.52  & $0.396\pm0.079$   & 0.978 & $0.379\pm0.176$   \\
        1.23 & $0.385\pm0.099$   & 1.526 & $0.342\pm0.070$  & 1.944 & $0.364\pm0.106$   &       &                  \\
        \bottomrule
    \end{tabular*}
\end{table*}

\begin{table*}[ht!]
    \centering
    \caption{$f(z)$ measurements from Ref.\,\cite{Sahlu:2024dxp} (left) and $\sigma_8$ measurements from Refs.\,\cite{Pezzotta:2016gbo,delaTorre:2016rxm,Shi:2017qpr} (right).}
    \label{tab:combined_f_sigma}
    
    \begin{minipage}[t]{0.48\textwidth}
        \centering
        \renewcommand{\arraystretch}{1.2}
        \begin{tabular}{cc}
            \toprule
            $z$    & $f(z)$ \\
            \midrule
            0.013 & $0.56 \pm 0.07$  \\
            0.10  & $0.464 \pm 0.04$ \\
            0.15  & $0.490 \pm 0.145$ \\
            0.18  & $0.49  \pm 0.12$ \\
            0.22  & $0.6   \pm 0.10$ \\
            0.35  & $0.7   \pm 0.18$ \\
            0.41  & $0.7   \pm 0.07$ \\
            0.55  & $0.75  \pm 0.18$ \\
            0.60  & $0.73  \pm 0.07$ \\
            0.60  & $0.93  \pm 0.22$ \\
            0.77  & $0.91  \pm 0.36$ \\
            1.40  & $0.99  \pm 0.19$ \\
            \bottomrule
        \end{tabular}
        \label{tab:f} 
    \end{minipage}
    \hfill 
    \begin{minipage}[t]{0.48\textwidth}
        \centering
        \renewcommand{\arraystretch}{1.2}
        \begin{tabular}{cc}
            \toprule
            $z$   & $\sigma_8$ \\
            \midrule
            0.10 & $0.769 \pm 0.105$ \\
            0.60 & $0.52  \pm 0.06$  \\
            0.86 & $0.48  \pm 0.04$  \\
            \bottomrule
        \end{tabular}
        \label{tab:sigma8}
    \end{minipage}
\end{table*}

\bibliographystyle{JHEP}
\bibliography{References.bib}

\end{document}